%% file: main.tex
\documentclass[%
    reprint,
    superscriptaddress,
    amsmath,
    amssymb,
    aps,
    floatfix,
]{revtex4-2}

\usepackage{graphicx}% Include figure files
\usepackage{dcolumn}% Align table columns on decimal point
\usepackage{bm}% bold math
\usepackage{siunitx} % for \unit and SI units
\usepackage{newfloat}
\DeclareFloatingEnvironment[name={Supplementary Figure}]{suppfigure}
\DeclareFloatingEnvironment[name={Supplementary Table}]{supptable}

\usepackage{hyperref}
\usepackage{xcolor}

\newcommand{\Sr}{\ensuremath{{}^{87}\text{Sr}}}
\newcommand{\Yb}{\ensuremath{{}^{171}\text{Yb}}}
\newcommand{\Al}{\ensuremath{{}^{27}\text{Al}^{+}}}
\newcommand{\bacon}{BACON21}
\newcommand{\AlSr}{\ensuremath{\text{Al}^{+}/\text{Sr}}}
\newcommand{\AlYb}{\ensuremath{\text{Al}^{+}/\text{Yb}}}
\newcommand{\YbSr}{\ensuremath{\text{Yb}/\text{Sr}}}

\begin{document}

    \title{Atomic clock frequency ratios with fractional uncertainty \texorpdfstring{$\leq 3.2 \times 10^{-18}$}{<=3.2E-18}}

    \date{\today}

\author{Alexander Aeppli}
\altaffiliation{Present address: Atom Computing, Boulder, CO, USA}
\affiliation{JILA, National Institute of Standards and Technology and the University of Colorado, Boulder, CO, USA}
\affiliation{Department of Physics, University of Colorado, Boulder, CO, USA}

\author{Willa J. Arthur-Dworschack}
\affiliation{Department of Physics, University of Colorado, Boulder, CO, USA}
\affiliation{Time and Frequency Division, National Institute of Standards and Technology, Boulder, CO, USA}

\author{Kyle Beloy}
\affiliation{Time and Frequency Division, National Institute of Standards and Technology, Boulder, CO, USA}

\author{Caitlin M. Berry}
\affiliation{Statistical Engineering Division, National Institute of Standards and Technology, Boulder, CO, USA}

\author{Tobias Bothwell}
\altaffiliation{Present address: Infleqtion, Boulder, CO, USA}
\affiliation{Time and Frequency Division, National Institute of Standards and Technology, Boulder, CO, USA}

\author{Angela Folz}
\affiliation{Department of Physics, University of Colorado, Boulder, CO, USA}
\affiliation{Associate, Statistical Engineering Division, National Institute of Standards and Technology}

\author{Tara M. Fortier}
\affiliation{Time and Frequency Division, National Institute of Standards and Technology, Boulder, CO, USA}

\author{Tanner Grogan}
\affiliation{Department of Physics, University of Colorado, Boulder, CO, USA}
\affiliation{Time and Frequency Division, National Institute of Standards and Technology, Boulder, CO, USA}

\author{Youssef S. Hassan}
\altaffiliation{Present address: Atom Computing, Boulder, CO, USA}
\affiliation{Department of Physics, University of Colorado, Boulder, CO, USA}
\affiliation{Time and Frequency Division, National Institute of Standards and Technology, Boulder, CO, USA}

\author{Zoey Z. Hu}
\affiliation{JILA, National Institute of Standards and Technology and the University of Colorado, Boulder, CO, USA}
\affiliation{Department of Physics, University of Colorado, Boulder, CO, USA}

\author{David B. Hume}
\affiliation{Department of Physics, University of Colorado, Boulder, CO, USA}
\affiliation{Time and Frequency Division, National Institute of Standards and Technology, Boulder, CO, USA}

\author{Benjamin D. Hunt}
\affiliation{Department of Physics, University of Colorado, Boulder, CO, USA}
\affiliation{Time and Frequency Division, National Institute of Standards and Technology, Boulder, CO, USA}

\author{Kyungtae Kim}
\email{kyungtae.kim@colorado.edu}
\affiliation{JILA, National Institute of Standards and Technology and the University of Colorado, Boulder, CO, USA}
\affiliation{Department of Physics, University of Colorado, Boulder, CO, USA}

\author{Amanda Koepke}
\affiliation{Statistical Engineering Division, National Institute of Standards and Technology, Boulder, CO, USA}

\author{Dahyeon Lee}
\affiliation{JILA, National Institute of Standards and Technology and the University of Colorado, Boulder, CO, USA}
\affiliation{Department of Physics, University of Colorado, Boulder, CO, USA}

\author{David R. Leibrandt}
\altaffiliation{Present address: Department of Physics and Astronomy, University of California, Los Angeles, CA, USA}
\affiliation{Department of Physics, University of Colorado, Boulder, CO, USA}
\affiliation{Time and Frequency Division, National Institute of Standards and Technology, Boulder, CO, USA}

\author{Ben Lewis}
\affiliation{JILA, National Institute of Standards and Technology and the University of Colorado, Boulder, CO, USA}
\affiliation{Department of Physics, University of Colorado, Boulder, CO, USA}

\author{Andrew D. Ludlow}
\affiliation{Department of Physics, University of Colorado, Boulder, CO, USA}
\affiliation{Time and Frequency Division, National Institute of Standards and Technology, Boulder, CO, USA}
\affiliation{Department of Electrical, Computer \& Energy Engineering, University of Colorado, Boulder, CO, USA}

\author{Mason C. Marshall}
\email{mason.marshall@nist.gov}
\affiliation{Time and Frequency Division, National Institute of Standards and Technology, Boulder, CO, USA}

\author{Nicholas V. Nardelli}
\altaffiliation{Present address: The University of Adelaide, Adelaide, South Australia, Australia}
\email{nicholas.nardelli@colorado.edu}
\affiliation{Department of Physics, University of Colorado, Boulder, CO, USA}
\affiliation{Time and Frequency Division, National Institute of Standards and Technology, Boulder, CO, USA}

\author{Harikesh Ranganath}
\affiliation{Time and Frequency Division, National Institute of Standards and Technology, Boulder, CO, USA}
\affiliation{Department of Electrical, Computer \& Energy Engineering, University of Colorado, Boulder, CO, USA}

\author{Daniel A. Rodriguez Castillo}
\affiliation{Department of Physics, University of Colorado, Boulder, CO, USA}
\affiliation{Time and Frequency Division, National Institute of Standards and Technology, Boulder, CO, USA}

\author{Jeffrey A. Sherman}
\affiliation{Time and Frequency Division, National Institute of Standards and Technology, Boulder, CO, USA}

\author{Jacob L. Siegel}
\email{jacob.siegel-1@colorado.edu}
\affiliation{Department of Physics, University of Colorado, Boulder, CO, USA}
\affiliation{Time and Frequency Division, National Institute of Standards and Technology, Boulder, CO, USA}

\author{Suzanne Thornton}
\affiliation{Statistical Engineering Division, National Institute of Standards and Technology, Boulder, CO, USA}
\affiliation{Department of Engineering, George Washington University, Washington, DC, USA}

\author{William Warfield}
\affiliation{JILA, National Institute of Standards and Technology and the University of Colorado, Boulder, CO, USA}
\affiliation{Department of Physics, University of Colorado, Boulder, CO, USA}

\author{Jun Ye}
\affiliation{JILA, National Institute of Standards and Technology and the University of Colorado, Boulder, CO, USA}
\affiliation{Department of Physics, University of Colorado, Boulder, CO, USA}

\collaboration{Boulder Atomic Clock Optical Network (BACON) Collaboration}
\input{figures.tex}

\begin{abstract}

We report high-precision frequency ratio measurements between optical atomic clocks based on \Al, \Yb, and \Sr. With total fractional uncertainties at or below $3.2 \times 10^{-18}$, these measurements meet an important milestone criterion for redefinition of the second in the International System of Units. Discrepancies in \Sr{} ratios at approximately $1\times10^{-16}$ and the \AlYb{} ratio at $1.6\times10^{-17}$ in fractional units compared to our previous measurements underscore the importance of repeated, high-precision comparisons by different laboratories. A key upgrade from our previous work is the use of a common ultrastable reference delivered to all clocks via a 3.6 km phase-stabilized fiber link between two institutions, enabling better accuracy and stability in frequency transfer. Derived from a cryogenic single-crystal silicon cavity, this reference improves comparison stability by a factor of 2 to 3 over previous systems, with an optical lattice clock ratio achieving a fractional instability of $1.3 \times 10^{-16}$ at 1 second.  Identifying individual clock stabilities with a three-corner hat measurement, we demonstrate the most stable optical lattice and trapped-ion clocks used in a multi-species comparison. By enabling faster comparisons, this stability will improve sensitivity to non-white noise processes and other underlying limits of state-of-the-art optical frequency standards.
    \end{abstract}

\maketitle

\textbf{\emph{Introduction.}} Frequency references based on spectroscopy of atomic transitions are the core of high-precision timekeeping, enabling global navigation satellite systems, network synchronization, and other technologies central to modern life.
The International System of Units (SI) defines the second based on a $9.2$~GHz hyperfine transition in Cs, with the best realizations of this definition having fractional systematic uncertainties at the $10^{-16}$ level~\cite{Weyers_2018,GerginovNISTF42025,BeattieCanadaCs2025}. Using frequencies at least $10^{4}$ times higher, optical atomic clocks achieve instability and systematic uncertainty two orders of magnitude lower than the best microwave standards~\cite{LudlowOpticalClockReview2015}. This motivates an effort to redefine the SI second in terms of one or more optical atomic clock frequencies~\cite{Dimarcq_2024}. High-precision frequency ratio measurements are a critical element of this effort, necessary to rigorously demonstrate optical clock performance and repeatability as well as optical frequency dissemination and timescale generation~\cite{schioppo_comparing_2022, grebingRealizationTimescaleAccurate2016, hachisuMonthslongRealtimeGeneration2018, yaoOpticalClockBasedTimeScale2019, milnerDemonstrationTimescaleBased2019}.
\par Beyond timekeeping, optical clock comparisons are among the most precise measurements \cite{bacon2021,HausserIndiumMultiIon2025}. This precision makes them well suited as probes for weak couplings introduced by physics beyond the standard model (BSM) \cite{SafronovaNewPhysicsReview2018}, with atomic clock ratios setting limits on potential BSM phenomena including various dark matter models \cite{bacon2021,KennedyDarkMatterClock2020,FilzingerClocksDM2023}, violations of local position invariance \cite{LangePositionInvariance2021}, and time variation of fundamental constants \cite{SafronovaFundamentalConstantVariation2019,SherrillFundamentalConstantVariation2023}.
Frequency comparisons can also enable relativistic geodesy, using frequency shifts between clocks to determine gravitational potential differences~\cite{Flury_relGeo_2016, grotti_chronometric_2024}.

We present frequency ratio measurements between three different optical frequency standards, all located in Boulder, Colorado: the \Al{} clock \cite{MarshallAlEvaluation2025} and \Yb{} optical lattice clock (OLC) \cite{mcgrew_Yb2018} located at the National Institute of Standards and Technology (NIST), and the \Sr{} OLC \cite{aeppli_SrAccuracy} located at JILA. These three optical clock species share similar electronic structures including narrow ($<10$~mHz), environmentally insensitive $^{1}S_{0}\rightarrow{}^{3}P_{0}$ clock transitions. A frequency comparison between these three species was previously completed in 2021~\cite{bacon2021} (\bacon{}). Since this measurement, the \Al{} and \Sr{} clocks and the frequency distribution network have been rebuilt and undergone new systematic evaluations~\cite{MarshallAlEvaluation2025,aeppli_SrAccuracy}. The \Yb{} clock operates largely as previously evaluated \cite{mcgrew_Yb2018}, with updated evaluations of certain systematic effects presented in the Supplemental Material \cite{SeeSupplementalMaterial}.

The optical local oscillators of all three clocks are phase-locked to the JILA cryogenic silicon laser reference cavity (Si cavity)~\cite{mateiMmLasersSub102017, oelker_Si2019} via a 3.6-km optical fiber network. This reduces comparison instability, enabling statistical uncertainty $<5\times10^{-18}$ for each frequency ratio in two hours of averaging time. We report the three frequency ratios with total fractional uncertainties $\leq 3.2\times10^{-18}$. In addition to being the most precise measurements of any physical quantity (other than symmetry tests, e.g. \cite{BressiNeutrality2011}), the ratios reported in this work meet the $5 \times 10^{-18}$ threshold identified in the roadmap for redefinition of the SI second~\cite{Dimarcq_2024, HausserIndiumMultiIon2025}.

\overviewfigure

\textbf{\emph{Optical network.}} 
Figure~\ref{fig:overview} shows a simplified schematic of the phase-coherent optical fiber network allowing frequency transfer between NIST and JILA~\cite{bran_documentation, Ye_BRAN_03}. $1.5$~\unit{\mu m} light stabilized to the Si cavity is transmitted to NIST via one fiber branch of the network. An optically encoded maser signal~\cite{narbonneauHighResolutionFrequency2006} is transferred from NIST to JILA through the same fiber with a wavelength division multiplexer, providing a common microwave reference between laboratories.

Each clock laser is pre-stabilized to local ultra-low expansion (ULE) glass cavities and then stabilized at low bandwidth ($<10$~kHz) to frequency combs locked to the Si-stabilized $1.5$~\unit{\mu m} light~\cite{Nardelli_2023,FortierTiSaph2006, oelker_Si2019,FortierCombReview2019}.
This process transfers the low instability of the Si cavity to the three clock lasers. This reduces the Dick effect noise on the OLC comparison instability, and yields longer atom-light coherence for longer spectroscopy times and reduced quantum projection noise (QPN) in the \Al{} clock~\cite{MarshallAlEvaluation2025}.

\ratioplot

\textbf{\emph{Measurement results.}} Between January 16 and March 21, 2025, all three clocks were compared nine times, and \Yb{} and \Sr{} OLCs were compared on four additional days. The results are presented in Fig.~\ref{fig:ratios}. In Fig.~\ref{fig:ratios}(a), we show the overlapping Allan deviation of the fractional ratios for a single day. A white frequency noise fit ($\sigma(\tau)= \sigma_0/\sqrt{\tau/\mathrm{s}}$ with the averaging time $\tau$) to times after 100 s (the servo attack time of the \Al{} clock) is plotted as a solid line for each ratio. The statistical uncertainty is obtained by extrapolating to the total measurement time for each day. Together with the mean value, this provides one data point for each day's bin, as plotted in Fig.~\ref{fig:ratios}(b).

The average of $\sigma(\tau)$ over the campaign provides an estimate of the instability. The average fractional instability for the frequency ratio between \Yb{} and \Sr{} (\YbSr{}) is $1.3 \times 10^{-16}/ \sqrt{\tau/\text{s}}$, which is limited by technical noise and represents more than a factor of 2 improvement over \bacon{}. The frequency ratios involving \Al{} with \Sr{} (\AlSr{}), and with \Yb{} (\AlYb{}) are dominated by single-ion QPN, with average fractional instability of $3.9 \times 10^{-16}/ \sqrt{\tau/\text{s}}$ for both, a factor of 3 improvement, corresponding to a nearly 10-fold reduction in measurement time (for a given statistical uncertainty) compared to \bacon{}.

The high stability of OLCs reveals excess scatter in \YbSr{} with the reduced chi-squared statistic $\chi^{2}_{red}=6.4$, suggesting uncharacterized shifts. While this statistic is comparable to that of \bacon{}, the between-day variability is substantially reduced from $10.8(4.2)\times10^{-18}$ in 2021 to $3.3(9)\times10^{-18}$ in the present case~\cite{SeeSupplementalMaterial}.  The \AlYb{} and \AlSr{} ratios show $\chi^{2}_{red}$ values of 1.3 and 1.2, respectively. 

We use a comprehensive Bayesian model~\cite{bacon2021} to aggregate data binned by measurement day, together with systematic uncertainties.  The frequency ratios and their standard uncertainties are:
\begin{equation}
\label{eq:ratios}
\begin{aligned}
    \AlSr{}=&2.611\ 701\ 431\ 781\ 462\ 7101(58), \\
    \AlYb{}=&2.162\ 887\ 127\ 516\ 663\ 6674(70), \\
    \YbSr{}=&1.207\ 507\ 039\ 343\ 337\ 7230(37).
\end{aligned}
\end{equation}
These represent fractional uncertainties of $2.2\times10^{-18}$, $3.2\times10^{-18}$ and $3.1\times10^{-18}$ for \AlSr{}, \AlYb{} and \YbSr{}, respectively. The Supplemental Material presents a detailed discussion of individual clock systematic uncertainties and the data analysis procedure, and also includes a simpler analysis with similar results~\cite{SeeSupplementalMaterial}. 

\figiii

\textbf{\emph{Discussion.}} All three ratios differ from the values measured in \bacon{} and the CIPM recommended values \cite{margolisCIPMListRecommended2024}, as shown in Fig.~\ref{fig:fig3}(a). 
The \AlYb{} ratio differs by $1.6\times10^{-17}$ or 2.4$\sigma$ from \bacon{}  in units of combined uncertainty $\sigma$.
The disagreement in the \Sr{} ratios is $14\sigma$, indicating a revision of the \Sr{} frequency by approximately $1 \times 10^{-16}$.
With recently improved evaluation of atomic coefficients, we could retroactively adjust the BACON21 results. The Sr and Yb frequency would increase by $7.5\times10^{-18}$ and $1.5\times10^{-18}$ respectively~\cite{aeppli_SrAccuracy, SeeSupplementalMaterial}, and this would increase the deviation from (14.5, 2.4, 13.9) to (15.4, 2.6, 14.7) in units of $\sigma$ for (\AlSr{}, \AlYb, \YbSr{}).

The \Sr{} clock was rebuilt with several improvements over the previous generation~\cite{bothwellJILASrIOptical2019}. The thermometers for black-body radiation (BBR) characterization are mounted on an in-vacuum translational stage, allowing calibration of the BBR shift before and after daily ratio data collection. The atomic density shift is corrected in-situ using local frequency imaging, and further suppressed by operating at a shallow optical lattice near the magic depth, where the collisional shift is minimized~\cite{aeppliHamiltonianEngineeringSpinorbitcoupled2022}. The lattice light shift and dc Stark shift were re-evaluated before and after the campaign to bound potential month-scale changes, and no significant drift was observed. 

The \Al{} clock was similarly rebuilt and improved over the previous generation~\cite{BrewerAlAccuracy2019}, with reduced excess micromotion and improved vacuum pressure. The stability of the Si cavity, transferred to the \Al{} clock laser, allows an extension of the probe time from 150 \unit{ms} to 1 second, requiring continuous Doppler cooling to control heating of the trapped ion's secular motion. For full details of each clock's operation, see the Supplemental Material~\cite{SeeSupplementalMaterial}.

We noticed the discrepancy with \bacon{} values in 2024 test runs and developed protocols to verify the accuracy of both the optical network and a number of systematic corrections for each clock. 
The JILA laser stability transfer system supports 10 different experiments, adding complexity in the network. For this campaign, we therefore implement a new loopback test scheme (see Fig.~\ref{fig:fig3}(b)) to evaluate the end-to-end accuracy and stability of the frequency network, including every element from Si cavity to \Sr{} atoms. The Si cavity-stabilized 1.5~\unit{\mu m} light is sent to NIST, where it references a frequency comb. The comb then stabilizes a $1397$~\unit{nm} laser, which is sent back to JILA in a parallel fiber. At JILA, this light is frequency doubled to $698$~\unit{nm} and compared against the \Sr{} clock laser, which is stabilized to a JILA comb referenced to the Si cavity, picked off just before the atoms. The resulting beat note indicates fractional instability $4.2\times 10^{-17}/\sqrt{\tau / \unit{s}}$, plotted as diamond points in Fig.~\ref{fig:ratios}(a), and a fractional frequency offset of $0.9(1.6)\times 10^{-19}$. Since the light travels through two 3.6-km fibers, we take this measurement as an upper bound on the network instability and inaccuracy.

To experimentally validate large systematic corrections on the atomic platforms, each clock in turn modulated certain systematic effects while the other two operated under nominal conditions. Systematic effects tested include the environment temperature, magnetic bias field magnitude, lattice depth, and spectroscopy time of the \Sr{} clock; the Doppler cooled motional energy of the \Al{} clock; and the magnetic bias field, measurement cycle time, and environment temperature of the \Yb{} clock. The Supplemental Material \cite{SeeSupplementalMaterial} includes discussion of each modulated systematic effect, as well as any significant differences in operational methodology for each clock between different days.

The network loopback measurement and the set of systematic checks developed in this campaign give us a high degree of confidence in the ratios measured here. The loopback measurement discussed above was significantly more complete in the current campaign, and largely rules out frequency transfer errors.  Between the systematic correction tests here and additional work reported elsewhere~\cite{BrewerMagconst2019, Siegel2025}, all known systematic effects that shift any of the three clocks by $1\times10^{-17}$ or more have been tested by direct clock comparisons. The applied corrections are confirmed at a level equal to or below the discrepancy in \Sr{} ratios. Additionally, the systematic effect evaluations for all three ratios have been improved~\cite{aeppli_SrAccuracy, MarshallAlEvaluation2025, SeeSupplementalMaterial}; the ratios also show improved between-day repeatability, in the form of reduced dark uncertainty.  Details are discussed in the Supplemental Material~\cite{SeeSupplementalMaterial}.  In summary, we believe that any origin for the discrepancy is more likely to have affected \bacon{} than this work.

More generally, we note that historical frequency ratio measurements shown in Fig.~\ref{fig:fig3}(a) (and extended in Fig.~\ref{fig:ratiohistoricaldataexpanded} of End Matter) exhibit disagreement at the few $\times10^{-16}$ level, albeit with significance at the 2-4 $\sigma$ level rather than 14. Additional repeated comparisons---both between JILA and NIST and at other laboratories---are necessary to properly resolve these discrepancies and set the stage for redefinition of the second. With a larger data set, sophisticated meta-analysis methods~\cite{KoepkeConsensusBuilder2017,ThompsonEllisonDarkUncertainty2011,BodnarHeterogeneity2020} can be employed to build consensus among measurements overscattered relative to their individual uncertainties.

\TCHcompressed

\textbf{\emph{Clock network analytics.}} Simultaneous comparison of three different clocks gives access to analytical tools not available for individual frequency-ratio measurements.
Comparing sets of multiple ratios, we are able to identify that the main discrepancy between comparison campaigns likely originates from a change in the \Sr{} frequency or optical network \footnote{Notably, the JILA-NIST link was tested differently in the two campaigns: in \bacon{}, a loopback test was performed only at $1.5~\mu$m, including the 3.6-km link and NIST frequency combs but omitting elements solely at JILA, whereas in the current campaign, the test extends to 698~nm as shown in Fig.~\ref{fig:fig3}(b).}.
Additionally, by restricting the dataset to only times when all three clocks are operating, we can analyze aspects of individual clocks and ratios, as shown in Fig.~\ref{fig:tchfig}.

Figure~\ref{fig:tchfig}(a) compares the normalized ratio results in pairs from each day of comparison. This demonstrates the degree of correlation in statistical fluctuations of the ratios, which helps understand individual clock and network behavior. On each subplot, one clock is common to both axes; the statistical uncertainty from that clock is therefore common. To determine the statistical uncertainties from each individual clock, we perform a three-cornered hat analysis \cite{GrayTCH1974}. As expected, the \AlYb{} and \AlSr{} ratios are strongly correlated, since statistical fluctuations in both ratios are dominated by common-mode statistical fluctuations due to QPN in the single-ion \Al{} clock. Conversely, the \AlYb{} and \YbSr{} ratios show no statistically significant correlation. 
Using three-cornered hat analysis on the nine-day concatenated data set, we derive stability estimates for each clock, plotted in Fig.~\ref{fig:tchfig}(b). We note that the overlapping Allan deviation is not well defined for gapped data in the presence of non-white noise, complicating this analysis for long averaging times \footnote{See for example the missing points from Fig.~\ref{fig:tchfig}(b) around $10^4$ s, where the model calculates an unphysical negative variance \cite{GrayTCH1974} and only an upper bound can be plotted.}.  While common-mode noise from the Si cavity means the three clocks are not statistically independent at short averaging times, at intermediate times from 100 to 5000 s the analysis gives a good approximation of the individual clock stability. Using a white noise model, we obtain fractional instabilities of $3.75(6)\times10^{-16}/\sqrt{\tau/\unit{s}}$ for \Al{}, $1.00(6)\times10^{-16}/\sqrt{\tau/\unit{s}}$ for \Yb{}, and $8.5(6)\times10^{-17}/\sqrt{\tau/\unit{s}}$ for \Sr{}.

The correlation of the \AlYb{} and \AlSr{} single-day ratios confirms that ratio statistical variation is dominated by single-ion QPN. However, these analytics are most useful when all three clocks have similar stability. Despite the significant improvement in \Al{} stability, the single-ion QPN is still too large to identify a source of between-day variability leading to the larger-than-expected scatter in \YbSr{} ratio values.  However, the \Al{} stability is sufficient to characterize the short-term stability of the lattice clocks -- including, to our knowledge, the first demonstration of sub-$10^{-16}/\sqrt{\tau / s}$ instability (\Sr{}) in an atomic clock with a three-corner-hat measurement. Future operation of a multi-ion \Al{} clock, extending the \Al{} probe time with differential spectroscopy \cite{KimDiffspec2023}, or extending the comparison averaging time will help identify the source of between-day variability. 

We additionally calculate Pearson correlation coefficients between the ratio means using their posterior distribution of the Bayesian model~\cite{margolisCIPMListRecommended2024, SeeSupplementalMaterial}.  These quantify the correlations between the final ratio values (Eqs.~\eqref{eq:ratios}). 
These include the systematic uncertainty of the shared clock, as well as correlations in statistical fluctuations.  
For pairs of ratios that share \Al{}, \Yb{}, and \Sr{}, these are 0.39, -0.74, and 0.27 respectively.

\textbf{\emph{Conclusion.}} We measure the frequency ratios between \Al{}, \Yb{}, and \Sr{} optical atomic clocks with fractional uncertainties $\leq 3.2\times10^{-18}$.
All three clocks re-evaluate systematic uncertainties and use a new distribution system for ultrastable light between institutes. 
In principle, this accuracy allows for chronometric leveling at the $3$~cm level. Additionally, simultaneous operation of more than two clocks gives access to valuable information about single-clock performance and network correlations.

Between this work and ratios we measured approximately 7 years earlier~\cite{bacon2021}, we observe a difference of approximately 14$\sigma$ in the ratios involving the \Sr{} clock, and a difference of approximately $2.4\sigma$ in the \AlYb{} ratio. Deliberate modulation of frequency-shifting systematic effects in each clock supports the accuracy evaluations at a level below these discrepancies. This time, the loopback measurement including both combs and the fiber link between NIST and JILA confirms accurate operation of all instruments common to the \AlSr{} and \YbSr{} ratios. Repeated measurements are necessary to resolve the ratio discrepancy and to strengthen confidence in results obtained by all laboratories.
The fast averaging time and validation protocols demonstrated in this work will enhance future measurements and ultimately help build consensus.  

\textbf{\emph{Acknowledgments.}}
We thank Franklyn Quinlan's group and Scott Papp's group for providing the 1.4~\unit{\mu m} laser and frequency doubler for the network loopback measurement. We are grateful to Christian Sanner and Alejandra Collopy for careful reading of the manuscript. 

We acknowledge support from the Office of Naval Research; the National Institute of Standards and Technology (including Award No. 70NANB18H006); the National Science Foundation Q-SEnSE Quantum Leap Challenge Institute (Grant No. OMA-2016244); the V. Bush Fellowship; the Sloan/Simons Foundation; and the National Science Foundation (Grant No. PHY-2317149).

The authors declare no competing interests.

This Letter is a contribution of the U.S. government, not subject to U.S. copyright.
\bibliographystyle{prsty}
\bibliography{bibliography}

\section{End Matter}
Figure~\ref{fig:ratiohistoricaldataexpanded} presents an expanded version of main text Fig.~\ref{fig:fig3}. Point types distinguish measurement methodologies: open symbols represent ratios between absolute frequency measurements relative to Cs standards; gray-shaded symbols indicate ratios via microwave link; colored symbols represent direct optical comparisons. This work, together with \bacon{} contributes the only direct measurements of the \Al{} ratios published to date. 

\ratiohistoricaldataexpanded

\clearpage

    \newpage
    \section{Supplement}

    \subsection{Reference values for the ratios}
    The 2021 CIPM reference frequency ratio values used in the main text are from Appendix B of Ref.~\cite{margolisCIPMListRecommended2024}.
    \begin{align*} 
    \AlSr{}  = 2.611\ 701\ 431\ 781\ 463\ 0187(390),\\
    \AlYb{} = 2.162\ 887\ 127\ 516\ 663\ 7047(240),\\
    \YbSr{} = 1.207\ 507\ 039\ 343\ 337\ 8451(160).
    \end{align*}

    \subsection{Discussion of gravitational redshift change}
    Changing gravitational redshifts are a potential source of ratio variation. National Geodetic Survey (NGS) markers installed in each clock lab \cite{NGSSurvey2019} are used to characterize gravitational redshifts between the clocks in both comparisons. While published data do not exist for Boulder, CO, nearby Denver is subsiding at an average rate of 2 mm/year \cite{OhenhenLandSubsidence2025}; this rate is more than an order of magnitude too low to explain a $10^{-16}$ discrepancy between measurements, even if it were completely differential between JILA and NIST.  While a more recent NGS survey \cite{NGSSurvey2025} did not re-measure the positions of the clock lab survey markers, it did confirm the height difference between NIST and JILA labs has changed by less than 5 mm - well below the ratio uncertainty.

    % \clearpage
\subsection{Data analysis}
The reported ratio values and uncertainties are obtained from individual-day ratio measurements generated as follows.  A time series of frequency values is recorded for each clock 
\footnote{Atomic transition frequency values for each clock are obtained by combining a measured beatnote $f_b$ against a comb tooth $n$ with the comb repetition rate $f_{rep}$, carrier envelope offset $f_0$, frequency multiplication factor $m$, and experimentally applied offsets $f_{os}$.  $f_{rep}$  is set by the comb lock to the Si cavity; $f_0$ is controlled to a set value; and $m$ and $f_{os}$ are set experimentally by each clock.  Together, the measured atomic transition frequency is $\nu=m\left(nf_{rep}+f_0+f_b\right)+f_{os}$ \cite{FortierCombReview2019}.  We note that all RF sources are referenced to a UTC(NIST) traceable maser and measured by frequency counters referenced to the same maser.  Because the absolute inaccuracy of the maser is small compared to the relevant optical frequencies and the counter resolution, we eliminate counter digitization noise by treating controlled RF frequencies as exact so long as the counted and set frequencies agree.}.
The frequencies are corrected for systematic shifts, linearly interpolated to a 1~s interval, and divided point-by-point to generate a time series of ratio values for each clock pair.  
The daily ratio value is the mean of this time series, while its statistical uncertainty is the Allan deviation extrapolated to the full measurement time and its systematic uncertainty is the quadrature sum of uncertainties from the two clocks involved.
To aggregate the daily data for each ratio into a final ratio value with total uncertainty, we use a comprehensive Bayesian model updated from \bacon{}~\cite{bacon2021} and summarized below.  We provide the results of two additional analysis methods in the subsequent section, for ease of comparison to previous clock frequency ratio measurements.
   
    \subsubsection{Comprehensive Bayesian model}
We obtain all reported parameter estimates and uncertainties using a hierarchical Bayesian model extended from \cite{bacon2021}. The Bayesian approach involves obtaining posterior distribution samples using Markov chain Monte Carlo and calculating summary statistics (i.e., mean, standard deviation, quantiles) to estimate the parameters in our scientifically motivated model. This model accounts for systematic effects, including shifts to each of the three clocks as well as shared geopotential and network effects. Excess daily scatter is treated as a clock-specific effect with zero mean, introducing further correlations into the ratio measurements. Daily statistical instabilities of, and statistical correlations between, ratio measurements are estimated from within-day data and input into the Bayesian model.

We expand the hierarchical Bayesian model used in \cite{bacon2021} to model all three ratios at once. For notational convenience, we use $X$ to denote the \AlSr{} ratio, $Y$ for \AlYb{}, and $Z$ for \YbSr{}. Lower case letters denote an observation of the ratio (e.g., $y_6$ is the observed \AlYb{} ratio on the sixth day). The likelihood for our Bayesian model is defined as follows. First, we model
\[    
    \begin{aligned}
\eta_X |\mu_X, \sigma_{G, X}, \sigma_{N, X} &\sim \mathcal{N}(\mu_X, \sigma_{G, X}^2 + \sigma_{N, X}^2) \\
\eta_Y |\mu_Y, \sigma_{G,Y}, \sigma_{N,Y} &\sim \mathcal{N}(\mu_Y, \sigma_{G, Y}^2 + \sigma_{N, Y}^2)\\
\eta_Z |\mu_Z, \sigma_{G, Z}, \sigma_{N, Z} &\sim \mathcal{N}(\mu_Z, \sigma_{G, Z}^2 + \sigma_{N, Z}^2). \\
\end{aligned}
\]
The vertical bar notation denotes a conditional distribution, where the variable on the left depends on the variables on the right.  $\mathcal{N}$ denotes a normal distribution. The $\mu$ represent the true, unknown values of the ratios which cannot directly be measured because of systematic effects. Instead we observe $\eta$, which is centered on the true value $\mu$ and incorporates uncertainties due to systematic effects (e.g., $\sigma_N, \sigma_G$, the network and geopotential effects, listed in Supplementary Tab.~\ref{tab:sysuncs}).  

    \begin{supptable}
    \renewcommand{\arraystretch}{1.5} 
    \caption{Fractional ratio uncertainties due to systematic effects (geopotential $\sigma_G$ and network $\sigma_N$). The values are identical to those used in~\cite{bacon2021}, with the exception of network uncertainties involving \Sr{} clock, which are updated based on the loopback measurement illustrated in the main text.}
    \label{tab:sysuncs}
\begin{tabular}{lcc} 
\hline
\hline
     & $\sigma_{G}$ ($\times 10^{-18}$) & $\sigma_{N}$ ($\times 10^{-18}$)\\ 
\hline
    \AlSr{} & 0.4 & 0.2 \\
    \AlYb{} & 0.2 & 0.3 \\
    \YbSr{} & 0.4 & 0.2 \\
    \hline
    \hline
    \end{tabular}
\end{supptable}

Then the observed ratios for each day $i$ can be modeled as
\[    
    \begin{aligned}
\begin{pmatrix}
x_i \\
y_i \\
z_i
\end{pmatrix}
\Bigg|&
\eta_X, \eta_Y, \eta_Z, \alpha, \beta, \gamma, \boldsymbol{a}, \boldsymbol{b}, \boldsymbol{c},\boldsymbol{\lambda_A},\boldsymbol{\lambda_B},\boldsymbol{\lambda_C}, \boldsymbol{\Sigma}
\sim \\
& \mathcal{N}\left(
\begin{pmatrix}
\eta_X + a_i\alpha - c_i\gamma +\lambda_{A,i}-\lambda_{C,i}\\
\eta_Y + a_i\alpha - b_i\beta +\lambda_{A,i}-\lambda_{B,i}\\
\eta_Z + b_i\beta - c_i\gamma +\lambda_{B,i}-\lambda_{C,i}
\end{pmatrix}
, \boldsymbol{\Sigma}_i
\right),
\end{aligned}
\]
Here, 
$\mathbf{a}, \mathbf{b}$ and $\mathbf{c}$ are observed daily systematic uncertainties for \Al{}, \Yb{} and \Sr{}, respectively, and $\alpha, \beta, \gamma \sim \mathcal{N}(0, 1)$. 
The product of these two quantities (e.g., $a_i \alpha$) models systematic shifts to each clock frequency. The $\lambda_{A, i} \sim \mathcal{N}(0, \xi_A^2)$, $\lambda_{B, i} \sim \mathcal{N}(0, \xi_B^2)$, and $\lambda_{C, i} \sim \mathcal{N}(0, \xi_C^2)$ represent random effects corresponding to between-day variability, where $\xi_{A}$, $\xi_B$, and $\xi_C$ are the ``dark uncertainty" attributable to \Al{}, \Yb{}, and \Sr{}, respectively~\cite{KoepkeConsensusBuilder2017}. These parameters quantify the amount of excess scatter on each clock's frequency due to unidentified sources. In addition, we take into account the correlation between the ratios arising from the common clock used in the ratios. The matrix
\[
{\Sigma}_{i} = 
\begin{pmatrix}
\sigma_{X,i}^2 & \sigma_{XY,i} & \sigma_{XZ,i} \\
\sigma_{XY,i} & \sigma_{Y,i}^2 & \sigma_{YZ,i} \\
\sigma_{XZ,i} & \sigma_{YZ,i} & \sigma_{Z,i}^2
\end{pmatrix}
\]
is the covariance matrix for the daily statistical fluctuations. The diagonal components $\sigma_{R, i}$ (where $R$ denotes the ratios) are obtained using the procedure described in Fig.~\ref{fig:ratios}(a). The covariances are estimated from the concatenated ratio values with their mean values subtracted for each day. We first compute the correlation coefficient $\rho(R, R')$ with data for the period where the ratio measurements overlap. Then, we weight each ratio's uncertainty by a factor that accounts for the daily overlap period $T_{overlap, i}$ and their total measurement periods $T_{R, i}$ and $T_{R', i}$. Explicitly, $\sigma_{RR', i}=\rho(R, R')\sigma_{R, i}\sigma_{R', i} T_{overlap, i}/\sqrt{T_{R, i} T_{R', i}}$~\cite{margolisGuidelinesEvaluationReporting2020a}.
 For the four days with only $z_i$ (see also Fig.~\ref{fig:ratios}(b)),  we use a marginalized form, i.e., $z_i|\eta_Z, \beta, \gamma, \boldsymbol{b}, \boldsymbol{c}, \boldsymbol{\lambda_B},\boldsymbol{\lambda_C},\boldsymbol{\Sigma} \sim \mathcal{N}(\eta_Z + b_i\beta - c_i\gamma +\lambda_{B,i}-\lambda_{C,i}, \sigma_{Z, i}^2)$.

The result of this analysis is shown in Table~\ref{tab:bayesianresults} and Supplementary Fig.~\ref{fig:bayesianfigure}. 
Note that the pairs of ratios show different degrees of statistical correlation because of the different systematic and statistical uncertainty characteristics of each clock.
Additionally, note that the shapes of the dark uncertainty posterior distributions are influenced by the 
constraint that this parameter be non-negative and by the relative scale between dark uncertainties $\xi$ and daily statistical uncertainties.

The prior distribution for each $\mu$ is a normal distribution with mean 0 and standard deviation $10^{-14}$, and the prior for each $\xi$ is a truncated-normal distribution with mean $5\times10^{-15}$ and standard deviation of $10^{-14}$. We find the result is not sensitive to the choice of prior distribution. 

    \bayesianfigure

    \begin{supptable}
    \renewcommand{\arraystretch}{1.5} 
    \caption{Results of the comprehensive Bayesian model for $\mu$ and $\xi$, expressed as fractional difference from the CIPM 2021 reference values. The uncertainties represent the standard deviations of the posterior distributions shown in Supplementary Fig.~\ref{fig:bayesianfigure}. The numerical values of the ratios are shown in main text Eqs.~\eqref{eq:ratios}.  For $\xi$, we include the 68~\% credible interval as well as standard deviation, because the posterior distributions are not symmetric.  
    All values are in fractional units of $10^{-18}$.
    }
    \label{tab:bayesianresults}
    \begin{tabular}{ccc}
    \hline
    \hline
    Quantity & Mean & Uncertainty \\
    \hline
    $\mu_{\AlSr{}}$ & -118.2 & 2.2 \\
    $\mu_{\AlYb{}}$ & -17.2  & 3.2 \\
    $\mu_{\YbSr{}}$ & -101.1 & 3.1 \\
    $\xi_{\Al{}}$ & 2.3 & 1.7~(0.6, 3.8) \\
    $\xi_{\Yb{}}$ & 2.0 & 1.2~(0.7, 3.2) \\
    $\xi_{\Sr{}}$ & 2.2 & 1.2~(0.9, 3.3) \\
    \hline
    \hline
    \end{tabular}
    \end{supptable}

\subsubsection{Other analyses}
For ease of comparison to other clock frequency ratio measurements, we include two additional analyses.  First, we perform a weighted statistics analysis which accounts for overscatter in the data by multiplying each day's statistical uncertainty by the Birge ratio $\sqrt{\chi^2_{red}}$.  Second, we perform an additive dark uncertainty analysis described by \cite{DorscherYbSrRatio2021}, which is the familiar Mandel-Paule estimate from meta-analysis literature \cite{mandel1970interlaboratory,langan2019comparison}.  This models an unknown varying systematic by adding a factor $\xi_{MP}$ in quadrature with each day's statistical uncertainty, where $\xi_{MP}$ is chosen such that $\chi_{red}^2=1$ for the modified dataset.  
Table~\ref{tab:analysiscomparison} compares the results of these statistical methods.
The three methods gave mutually consistent results at $\leq 1.2\times10^{-18}$ for fractional mean values. This difference is mainly from the multivariate nature of the current model, which is absent in the other two models. The multivariate model accounts for correlations inherent in three simultaneous clock measurements, better informing all model parameter estimates. 

Both the Birge ratio method and the Mandel-Paule method include an adjustment factor to account for excess scatter in the data ($\sqrt{\chi^2_{red}}$ for the weighted statistics analysis and $\xi_{MP}$ for the Mandel-Paule method), whose value is determined by $\chi_{red}^2$ of the original dataset.  The calculation of the adjustment factor leads to important limitations for both of these methods.  First, as method-of-moments-based estimators using just summaries of the data (the mean and variance), the $\chi^2$-based estimates of the adjustment factor are statistically sub-optimal (i.e., not efficient) compared to a maximum likelihood estimate or an estimate from a Bayesian model.  Method-of-moments-based estimators are known to perform poorly (higher variance, lower accuracy) when the sample size is small, a common scenario for clock ratio data \cite{langan2019comparison}. Second, these methods do not account for uncertainty in the adjustment factor itself, which leads to overly optimistic (i.e., too small) uncertainty evaluations \cite{merkatas2019shades,possolo2023tracking}.  

Additionally, in cases where the overscatter is large compared to the individual data point statistical uncertainties (as is true for the Yb/Sr ratio here), the Birge ratio method can give a biased estimate of the mean value. By preserving the relative weights of the individual data points, it will over-weight points with low statistical uncertainty, even though they may not provide more information about the underlying distribution.
 
The hierarchical Bayesian model does not suffer from any of these limitations: it achieves optimal estimation and propagates all the recognized uncertainties accurately. In addition, it can model correlations between the ratios being aggregated. We note that all models here incorporating dark uncertainty or between-day variability assume that this variability affects every measurement equally and that the measurements themselves are free from any unknown systematic bias.  Future work will involve a rigorous statistical comparison of methods currently used in the clock metrology literature, detailing model assumptions and the circumstances under which they are, or are not, appropriate.
\begin{supptable}
\renewcommand{\arraystretch}{1.5} 
\caption{Comparison of statistical analysis methods. $\mu_B$ and $\sigma_B$ are the mean value and uncertainty of the ratio based on the Birge ratio method. $\mu_{MP}$ and $\sigma_{MP}$ are the mean value and uncertainty of the ratio from the Mandel-Paule method, while $\xi_{MP, a}$ is the dark uncertainty estimate for the ratio $a$. All values are in fractional units of $10^{-18}$, with means referenced to the mean value from the Bayesian analysis.}
\label{tab:analysiscomparison}
\begin{tabular}{lccccc}
\hline
\hline
Ratio & $\mu_B$ & $\pm~\sigma_B~~$ & $\mu_{MP}$ & $\pm~\sigma_{MP}~~$ & $\xi_{MP, a}$ \\
\hline
Al$^{+}$/Sr~~ & -0.7 & $\pm$ 1.9~~ & -0.8 & $\pm$ 2.0 ~~& 1.7 \\
Al$^{+}$/Yb~~ & 1.1 & $\pm$ 3.0 ~~& 1.2 & $\pm$ 3.0  ~~& 1.5 \\
Yb/Sr ~~      & -0.5 & $\pm$ 3.0 ~~& -0.7 & $\pm$ 3.0  ~~& 2.9 \\
\hline
\hline
\end{tabular}
\end{supptable}

\input{schedule}
\input{sr_supplement}
\input{supplement_al}

\input{Yb_Supplement}

\subsection{Full network architecture}
In Fig.~\ref{fig:fullnetworkschematic}, we present a more detailed diagram of the network architecture, including all frequency offsets.
\fullnetworkschematics
\clearpage

\end{document}

%% file: figures.tex
%  Figure 1
\newcommand{\overviewfigure}{
\begin{figure}[!t]
    \includegraphics[width=\columnwidth]{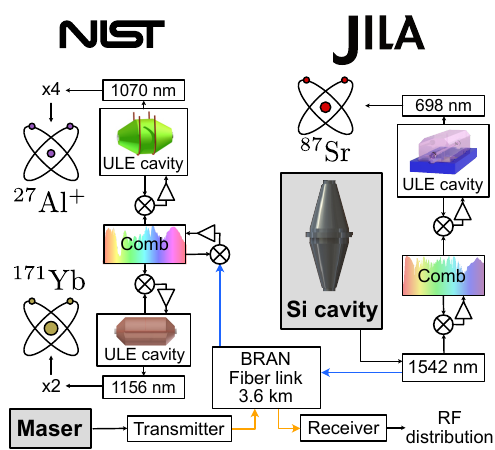}
    \caption{Simplified schematic of the Boulder optical clock network. Two independent oscillators---a cyrogenic silicon (Si) optical cavity and a hydrogen maser for the radio-frequency---are distributed across multiple laboratories via the 3.6 km-long Boulder Research and Administration Network (BRAN) optical fiber link. The \Al{} and \Yb{} clocks are located at NIST, and the \Sr{} clock is located at JILA. Local ULE cavity-referenced lasers in all laboratories are steered by a common 1542~nm laser locked to the Si cavity at JILA. Frequency combs at JILA and NIST transfer the Si cavity's stability to the clock wavelengths. A stable microwave signal from a maser at NIST is distributed optically to JILA. 
     \label{fig:overview}}
\end{figure}
}

% Figure 2
\newcommand{\ratioplot}{
\begin{figure*}[!th]
    \includegraphics[]{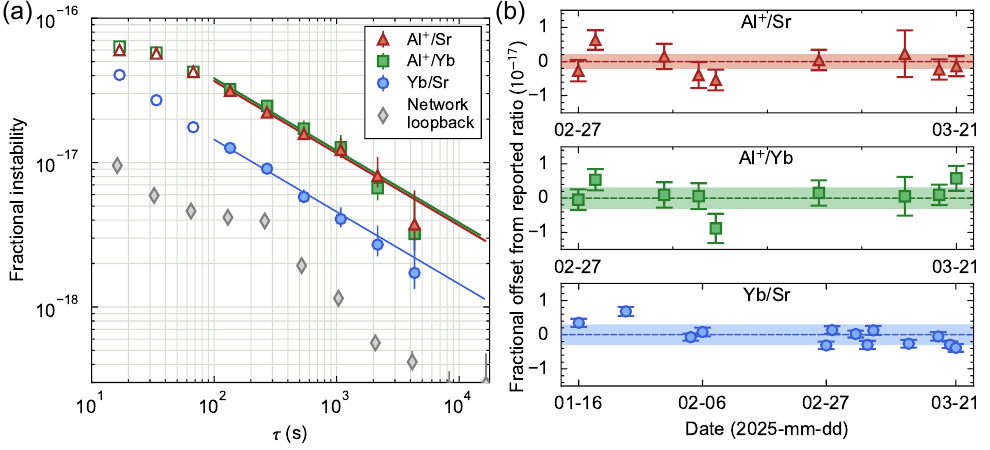}
    \caption{\label{fig:ratios} 
    Frequency ratio measurement results.
    \textbf{(a)} Fractional instability of the ratio measurements. The points are the overlapping Allan deviation of each fractional ratio, calculated for different averaging times $\tau$ and using concatenated data from a single measurement day. The solid lines are a white-frequency-noise model fit to points with $\tau$ greater than the servo attack time of $\sim100$~\unit{s} (color filled points). We extrapolate this fit to the total measurement time (the end of the fitted line) to obtain the statistical uncertainty.
    The gray diamonds are an upper bound on the instability of the fiber link, estimated by a loopback measurement. 
    \textbf{(b)} Daily ratio measurements are shown as fractional deviations from the final reported values (Eqs.~\eqref{eq:ratios}). The error bars represent statistical uncertainty obtained by the procedure described in \textbf{(a)}.  
    The colored shaded regions show the final fractional uncertainties including systematic uncertainties and are $ 2.2\times 10^{-18}$, $ 3.2\times 10^{-18}$, and $ 3.1\times 10^{-18}$ for \AlSr{}, \AlYb{}, and \YbSr{} ratios, respectively. 
    All uncertainties represent the 68 \% confidence interval. }

\end{figure*}
} 

% Fgiure 3
\newcommand{\figiii}{
\begin{figure}[tbp!]
    \begin{center}
        \includegraphics[width=\columnwidth]{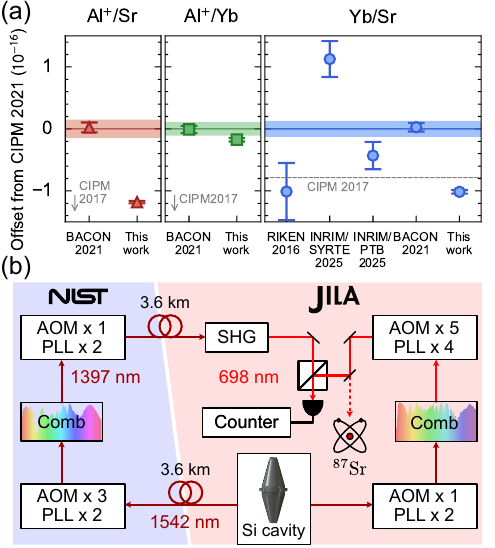}
        \caption{\textbf{(a)} Reported frequency ratio measurements with direct links~\cite{NemitzYbSrRatio2016,LindvallClockComparisons2025,bacon2021}. Only the five lowest uncertainty results are shown for \YbSr{}. The solid lines with shaded regions are the CIPM 2021 recommended values and their 1-$\sigma$ uncertainties.
        An expanded version is presented in End Matter Fig.~\ref{fig:ratiohistoricaldataexpanded}.
        \textbf{(b)} Schematic of the end-to-end loopback test for evaluating frequency network accuracy and stability. PLL: optical phase-locked loop. SHG: second harmonic generation. The acousto-optic modulators (AOMs) and PLLs use frequencies referenced to the maser, and are indicated for each path.
        }  
        \label{fig:fig3}
    \end{center}
\end{figure}
}

% Figure 4
\newcommand{\TCHcompressed}{
\begin{figure}[tbp!]
    \begin{center}
        \includegraphics[width=\columnwidth]{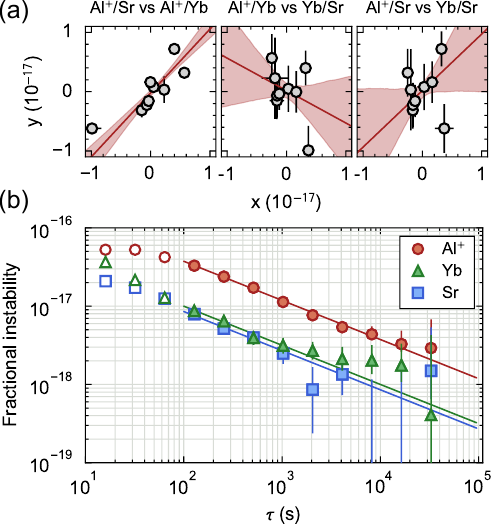}
        \caption{Insights making use of three clocks running simultaneously. \textbf{(a)} Correlations between day-to-day fractional variation in pairs of frequency ratios (y vs x). Each data point is one measurement day, restricted to times when all three clocks operated simultaneously. The mean of all nine days is subtracted from each data point. The \AlYb{} and \AlSr{} ratios are strongly correlated, as their fluctuations are dominated by single-ion quantum projection noise. The \YbSr{} ratio shows no statistically significant correlation with either of the ratios involving \Al{}.  
        Red lines and shaded areas are a linear fit to the data and 1-$\sigma$ uncertainty, with slope given by $1.08(18)$, $-0.55(76)$, and $0.95(97)$ for the three subplots. 
        \textbf{(b)} Estimating the fractional frequency stability of the three individual clocks. Points show overlapping Allan deviations for each clock, calculated using a three-cornered hat analysis. Solid lines are weighted white noise model fits to data points with averaging time $\tau> 100~\unit{s}$ (filled markers).
        }
        \label{fig:tchfig}
    \end{center}
\end{figure}
}

%Figure 5 for End Matter
\newcommand{\ratiohistoricaldataexpanded}{
\begin{figure*}[tbp!]
    \begin{center}
        \includegraphics[]{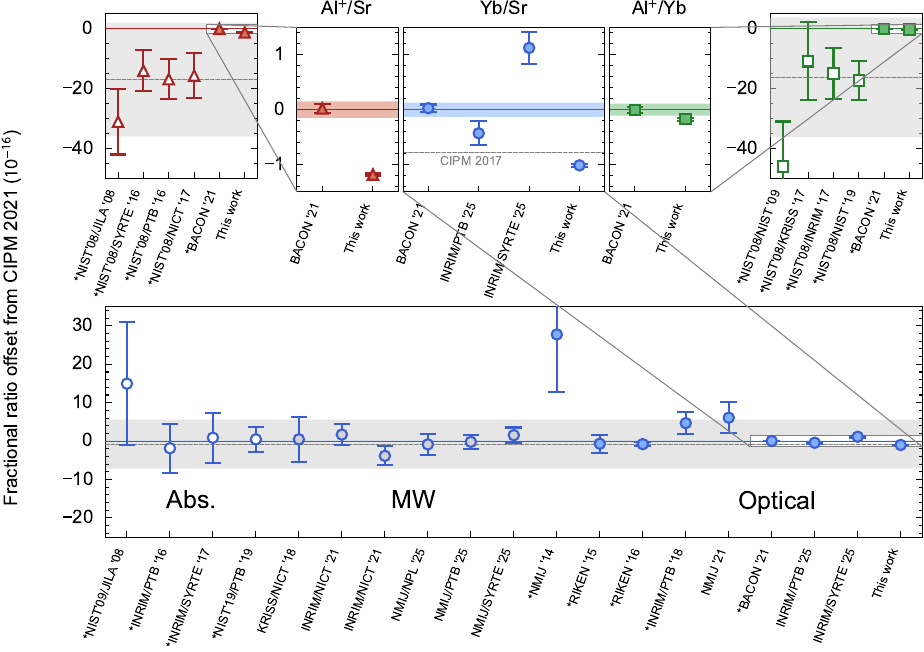}
        \caption{
        Evolution of frequency ratio measurements. Asterisks (*) indicate the values we presented in \bacon{}~\cite{bacon2021, akamatsuFrequencyRatioMeasurement2014, campbellAbsoluteFrequencyThe87Sr2008, grebingRealizationTimescaleAccurate2016, grottiGeodesyMetrologyTransportable2018, hachisuSItraceableMeasurementOptical2017a, kimImprovedAbsoluteFrequency2017a, lemkeSpin122009, lodewyckOpticalMicrowaveClock2016, mcgrewOpticalSecondVerifying2019, NemitzYbSrRatio2016, pizzocaroAbsoluteFrequencyMeasurement2017, rosenbandFrequencyRatioAl+and2008, takamotoFrequencyRatiosSr2015}. The errorbars represent 1-$\sigma$ uncertainty. Direct ratio measurements performed after CIPM 2017 are shown, organized by measurement method and year~\cite{fujiedaAdvancedSatelliteBasedFrequency2018, hisaiImprovedFrequencyRatio2021, pizzocaroIntercontinentalComparisonOptical2021, LindvallClockComparisons2025}. Colored lines and shaded regions represent the mean values and uncertainties from CIPM 2021~\cite{margolisCIPMListRecommended2024}, while gray lines and shaded regions indicate the mean ratio and uncertainty from CIPM 2017~\cite{riehleCIPMListRecommended2018a}. Marker styles denote different measurement methods: filled circles for direct optical measurements, open circles for absolute frequency measurements, and gray-filled circles for microwave measurements.}  
        \label{fig:ratiohistoricaldataexpanded}
    \end{center}
\end{figure*}
}

\newcommand{\fullnetworkschematics}{
\begin{suppfigure*}
    \includegraphics[width=0.8\textwidth]{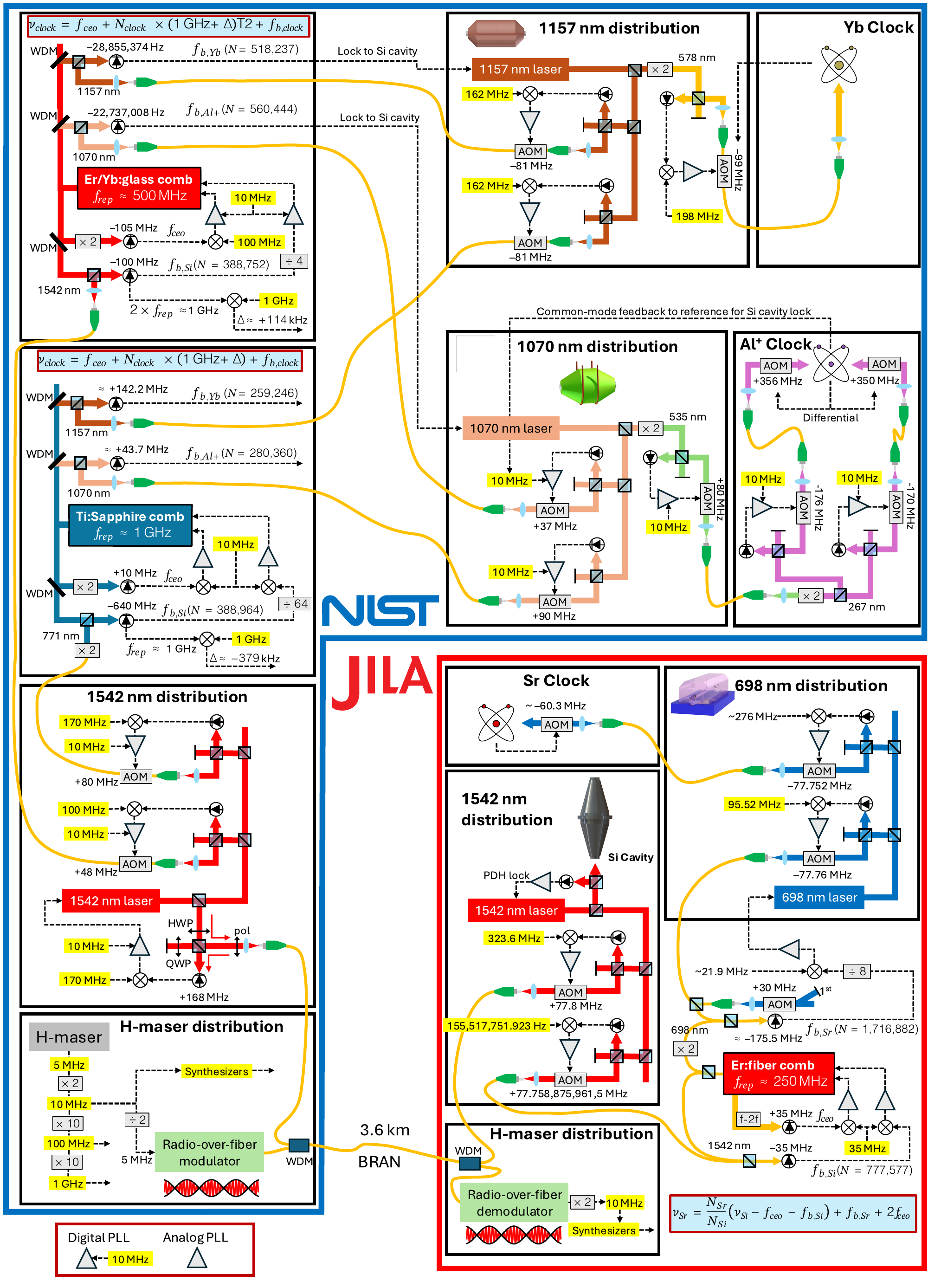}
    \caption{Full network layout. A Si-cavity-stabilized 1542 nm laser is sent to NIST, where it is amplified via a phase-locked loop with another laser. This laser is then distributed to two frequency combs, each locked to 1542 nm. Local ULE-cavity-stabilized lasers from the \Yb{} and \Al{} clock laboratories are sent to each comb. Beat notes from the Er/Yb:glass comb are sent to each clock laboratory to stabilize local oscillators to the Si cavity. A Ti:sapphire comb provides redundant monitoring of these frequency chains. A hydrogen maser (H-maser) signal distributed across NIST is transferred to amplitude modulations on the light sent from NIST to JILA~\cite{narbonneauHighResolutionFrequency2006}, and this signal is then distributed across the JILA laboratory.
    \label{fig:fullnetworkschematic}}
\end{suppfigure*}
}

\newcommand{\bayesianfigure}{
\begin{suppfigure}[h]
    \begin{center}
        \includegraphics[]{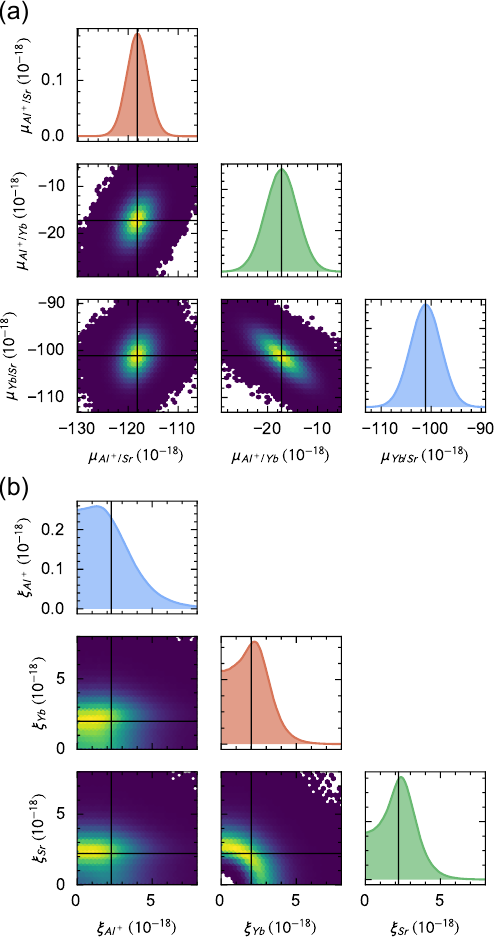}
        \caption{
        \textbf{(a)} The posterior distributions of ratio estimates $\mu$ and their bivariate histograms. The solid lines indicate the mean of the distribution. The Pearson correlation coefficients $r(R, R')$ are $r(\AlSr{}, \AlYb{}) = 0.39$, $r(\AlSr{}, \YbSr{}) = 0.27$, and $r(\AlYb{}, \YbSr{}) = -0.74$
        \textbf{(b)} The posterior distributions of dark uncertainties $\xi$ and their bivariate histograms. The solid lines indicate the mean of the distributions. The result is summarized in Table~\ref{tab:bayesianresults}. 
        The units are fractional, with the CIPM 2021 values as the reference.
        }
        \label{fig:bayesianfigure}
    \end{center}
\end{suppfigure}

}

%% file: schedule.tex
%%%%%
\subsection{Comparison Schedule}
Over 13 days between January 16 and March 21, 2025 (60691--60755 in Modified Julian Date), the \Sr{} optical lattice clock at JILA was compared with the \Yb{} optical lattice clock at NIST, with the \Al{} single-ion clock joining for the final 9 days. 
All clocks operated a majority of the time in a ``high accuracy'' configuration, and for a few days, each clock modulated systematic effects to explore the validity of applied systematic corrections. 
A comparison schedule is presented in Tab.~\ref{tab:comparison_schedule}, with ``nominal'' conditions indicating the clock was operating in a standard or near standard configuration. 

\begin{supptable}[!th]
    \caption[2025 Comparison Schedule]{Schedule of the comparison days during early 2025.    
    \label{tab:comparison_schedule} }
    \renewcommand{\arraystretch}{1.2} 
    % \begin{tabular} {\textwidth}{l x x x} 
    \centering
    \begin{tabular}{>{\centering\arraybackslash}m{1.5cm} 
                    >{\centering\arraybackslash}m{2.1cm} 
                    >{\centering\arraybackslash}m{2.1cm} 
                    >{\centering\arraybackslash}m{2.1cm}}
        \hline Day & \Sr{}  & \Al{} 
        & \Yb{}  \\
        \hline Jan. 16 & Nominal & N/A & Nominal \\
        Jan. 24 & Nominal & N/A & Nominal \\
        Feb. 04 & Nominal & N/A & Nominal \\
        Feb. 06 & Nominal & N/A & Nominal \\
        Feb. 27 &  \parbox[c][3em][c]{2cm}{\centering $370$~$^{\circ}$C oven temp. } & Nominal &
        Nominal \\
        Feb. 28 & \parbox[c][3em][c]{2cm}{\centering$T_{\text{rabi}}$=420~ms, 1 G bias} & Nominal & Nominal \\
        Mar. 04 &  \parbox[c][3em][c]{2cm}{\centering$19.6$~$^{\circ}$C  system temp. } & Nominal &
        Nominal \\
        Mar. 06 & $U = 23$~$E_{r}$ & Nominal & Nominal \\
        Mar. 07 & Nominal & Nominal & Nominal \\
        Mar. 13 & Nominal & \parbox[c][3em][c]{2cm}{\centering High Doppler temp.} & Nominal \\
        % Mar. 14 & Nominal & Comb issue & Comb issue \\
        Mar. 18 & Nominal & Nominal & \parbox[c][3em][c]{2cm}{\centering$T_{dead}$ extra 200~ms } \\
        Mar. 20 & Nominal & Nominal & High bias \\
        Mar. 21 & Nominal & Nominal & Nominal \\
        \hline
    \end{tabular}
\end{supptable}

%% file: sr_supplement.tex
\subsection{\Sr{} OLC operation}
\emph{\textbf{Improved design.}} The JILA strontium optical lattice clock and relevant systematics have been described extensively in Refs.~\cite{aeppli_SrAccuracy, kimEvaluationLatticeLight2023, aeppliHamiltonianEngineeringSpinorbitcoupled2022, bothwell2022}.
Rebuilt in 2020, this system differs significantly from the system used in the previous \bacon{} comparison and evaluated in Ref.~\cite{bothwellJILASrIOptical2019}.
The new clock incorporates an in-vacuum buildup cavity for the one-dimensional optical lattice, imaging spectroscopy, extensive active thermal control, and a retractable temperature probe for repeated measurements of the in-vacuum radiant environment.
The systematic shifts were recently evaluated in Ref.~\cite{aeppli_SrAccuracy}.
Key to this low systematic uncertainty is atomic confinement in a shallow optical lattice at the ``magic depth,'' eliminating the density shift~\cite{aeppliHamiltonianEngineeringSpinorbitcoupled2022}.
The in-vacuum cavity provides a large mode area lattice with stable lattice orientation, polarization, and intensity.

We note that the systematic effect correction differs by $7.3\times10^{-18}$ between \bacon{} and the current measurement, due to the change in the atomic coefficient for the dynamic BBR correction. The change comes from the updated measurement of an Einstein A-coefficient, data-driven modeling, and the improved computational accuracy~\cite{lisdatBlackbodyRadiationShift2021, aeppli_SrAccuracy}.

\emph{\textbf{Operational conditions.}} The Sr optical lattice clock at JILA was compared with NIST \Al{} and \Yb{} optical clocks, modulating systematics for four comparison days as noted in Tab.~\ref{tab:comparison_schedule}. 
The nominal Sr clock conditions are similar to Ref.~\cite{aeppli_SrAccuracy}: lattice depth $U = 11$~lattice photon recoil energies~($E_{r}$), $0.5$~G magnetic bias field; $22.1$~$^{\circ}$C environment temperature; $3 \times 10^{4}$ atoms; $400$~$^{\circ}$C oven temperature; Rabi spectroscopy pulse time $T_{rabi} = 1.0$~s (some measurements with $T_{rabi}= 1.3$~s), ground band cooled, single band selected (we reduce the lattice depth to select only a single motional band) sample with a temperature of $100$~nK along the weakly confined direction; $\nu_{lat}= \nu_{magic}+ 10$~MHz; $60$~s vacuum lifetime.
%%%%%%%
\par \emph{\textbf{Daily operation}}. 
To ensure a homogeneous environment and reduce thermal transients, we leave the system running the night before each comparison day. 
Daily operation begins with a measurement of the radiant temperature using the in-vacuum temperature probe described in Ref.~\cite{aeppli_SrAccuracy}.
We determine the temperature at the atom location as well as in the probe retracted position to bound the immersion error---in all cases $<10$~mK. 
To maintain the precision of the density shift and lattice light shift corrections, consistent laser cooling performance is essential.
We compensate the ambient magnetic field to the $1$~mG level on a daily basis, and lock the cooling lasers to the Si cavity (similar to the clock laser architecture shown in Fig.~\ref{fig:overview}).
The atomic temperature is measured using motional spectroscopy~\cite{blatt_Rabi_2009}. 
We also ensure coherent Rabi spectroscopy, confirming daily that the $1$~s contrast is roughly $95~\%$.
During operation, we make use of in-situ imaging of frequency gradients for point-by-point measurement and removal of residual density shifts~\cite{aeppliHamiltonianEngineeringSpinorbitcoupled2022}.
We have extensively evaluated the efficacy of this correction technique and have observed in all cases correction uncertainties $<10~\%$ of the total correction.
We use a $10~\%$ error on the magnitude of this correction as a conservative bound on this systematic effect. 
At the end of each comparison day, we again measure the radiant temperature in the chamber center; we observe temperature drift $<10$~mK on all days.

\begin{suppfigure}[!t]
    \includegraphics[width=\columnwidth]{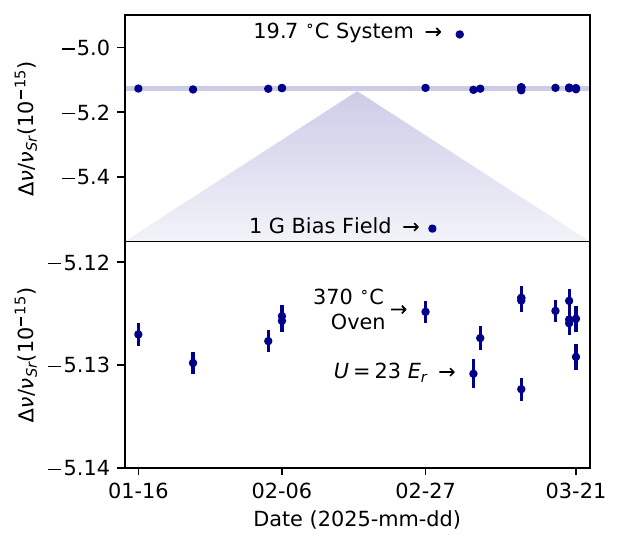}
    \caption{\label{fig:sr_corrections} 
    Daily \Sr{} frequency shifts in fractional frequency units $\Delta \nu/\nu_{Sr}$. 
    The top row presents frequency shifts for daily measurement, with the light blue shaded region highlighting the nominal operational conditions. 
    On some days, there are multiple data points separated by dead time..
    On the bottom, we zoom in to the data with the nominal operational condition (the shade on the top). 
    The error bars are the total systematic uncertainty including average point by point correction uncertainties and static correction uncertainties. 
    The four days with modulated Sr systematic effects as shown in Tab.~\ref{tab:comparison_schedule} are indicted on the plot.}
\end{suppfigure}

\begin{supptable}[!t]
\caption{\label{tab:Sr_ErrorBudget}%
Average shifts and uncertainties in the JILA 1D Sr OLC for standard operation during the clock comparison.
The * indicates shifts that are evaluated on a point-by-point basis, reported as the average value across all nominal runs.
The ${}^{\dagger}$ indicates the systematic is evaluated daily.
Shifts in bold are modulated during the comparison.
The redshift is reported to NGS marker S1B60V1.}
\begin{tabular}{lcc}
\hline \hline
         Effect &  Shift ($10^{-19}$) &  Uncertainty ($10^{-19}$)  \\
\hline
    \textbf{BBR}$^{\dagger}$ &   -49814.9 &   9.5 \\
    \textbf{Lattice light}$^{\dagger}$ & 10.3 &   4.9 \\
    \textbf{Density}* &   -18.8 &    1.9 \\
    \textbf{Quad. Zeeman}* &   -1388.3 &   1.6 \\
    Background gas &   -4.7 &    0.5 \\
    DC Stark & -0.04 &  0.03 \\ \hline
    Total & -51216.5 & 11.0 \\ \hline \hline
    Redshift to marker & -85.6 & 3.5 \\ 
    Total with redshift & -51302.1 & 11.5 \\

\hline \hline
\end{tabular}
\end{supptable}
\par \emph{\textbf{Systematics modulation.}} The total shifts for each comparison day are plotted in Fig.~\ref{fig:sr_corrections}, with the top plot containing all points and the bottom plot illustrating a narrow range around the nominal configuration.
The days with modulated systematics are indicated on the plots.
In Tab.~\ref{tab:Sr_ErrorBudget}, we report the average shifts and shift uncertainties for all applied corrections.
Minor shifts, such as shifts due to the probe, Doppler motion, first order Zeeman, and tunneling, are all $<10^{-19}$ and do not contribute appreciably to the total shift sum or uncertainty.
The average total uncertainty is slightly higher than the uncertainty reported in Ref.~\cite{aeppli_SrAccuracy} primarily due to the inclusion of the gravitational redshift and slightly higher temperature uncertainty during some measurement days.
%%%%
\par The strontium oven is both a BBR emitter and a system heat load, indirectly modifying the BBR environment. 
The oven is part of a commercial atomic beam source and has no direct line-of-sight to the atoms.
We have measured no radiant coupling between the oven and the in-vacuum temperature sensor.
On February 27, we operate with an oven at $370$~$^{\circ}$C, $30$~$^{\circ}$C colder than the nominal configuration.
We observe no difference in the ratio measurements on this day.
\par We use the least magnetically sensitive magnetic sublevel transition $|{}^1S_0, \, m_F = \pm 5/2 \rangle \rightarrow |{}^3P_0, \, m_F = \pm 3/2 \rangle$, alternating magnetic sublevel signs during the clock measurement and taking the transition frequency average to reject the first order Zeeman shift.  
The second order Zeeman correction is applied in a point by point fashion using the measured splitting between magnetic sublevel clock transitions, $\Delta_{m_F}$. 
The vector shift is calculated using this splitting and previously measured lattice parameters~\cite{shi_polarizabilities_2015}.
Under standard operation, $\Delta_{m_F} = -22$~Hz resulting in a second order Zeeman shift of $-1.388 \times 10^{-16}$.
On February 28, 2025, we use a bias field roughly double the standard field strength such that $\Delta_{m_F} = -44.7$~Hz, giving a second order Zeeman shift of $-5.720 \times 10^{-16}$. 
During the same comparison measurement, we use a shorter $420$~ms Rabi time.
The clock ratios on February 28, 2024 are consistent with other days despite generating a $4 \times 10^{-16}$ shift difference, so which gives confidence that the second order Zeeman shift is properly corrected.
\par The environment temperature is typically $22.1$~$^{\circ}$C with an average measurement uncertainty of 8~mK. 
On March 4, 2025, we reduced the system temperature to $19.6$~$^{\circ}$C, indicated in the top plot of Fig.~\ref{fig:sr_corrections}. 
The average BBR shift under nominal conditions is near $-4.981 \times 10^{-15}$, accounting for a majority of the applied correction. 
The temperature modulation leads to a shift difference of approximately $2 \times 10^{-16}$.
Since ratios agree $<4 \times 10^{-18}$, this measurement is a direct characterization of the BBR correction at roughly one part in 50. 
\par The first lattice light shift evaluation was completed in 2022 and published in 2023~\cite{kimEvaluationLatticeLight2023}.
To ensure robustness in the lattice light shift measurement, we repeat light shift measurements before the comparison in July 2024 and after the comparison in May 2025. 
Both evaluations agree within uncertainty, and we use a fit to the combined data to extract relevant light shift parameters~\cite{ushijima_ls_2018}.
Under nominal operation during the comparison, our light shift is $(1.0 \pm 0.5) \times 10^{-18}$ with $U = 11E_r$ and the detuning from the magic wavelength $\delta = 10$~MHz.
On March 6, 2025, we operate with $U = 23E_r$, leading to a light shift of $(-1.3 \pm 0.8) \times 10^{-18}$.
This higher depth also leads to a larger density shift.
Nevertheless, no difference in ratios is observed during this campaign.
%%%
\par \emph{\textbf{Gravitational redshift.}} A 2019 survey determined the gravitational redshift difference between the JILA and NIST, and installed three permanent markers near the Sr clock~\cite{NGSSurvey2019}.
We evaluate the height difference between the Sr atom location and one marker, S1B60V1.
We begin by determining the height of the blue MOT with respect to the in-vacuum temperature probe using a camera and known dimensions of the probe. 
We then align a laser level to the probe and project a level line to a structure around the optical table.
We fill a plastic tube with water to this marker level, and take the other end to the north wall of the lab to project the potential level near the marker.
Finally, we measure the height of the water surface with respect to the survey marker using a calibrated ruler.
With respect to the S1B60V1 marker, we find an atomic height of $-7.86 \pm 0.32$~cm, corresponding with a redshift of $-(8.56 \pm 0.35) \times 10^{-18}$.

% \bibliography{bibliography}
% \end{document}

%% file: supplement_al.tex
\renewcommand{\thefigure}{S\arabic{figure}}\makeatother
\renewcommand{\thetable}{S\arabic{table}}\makeatother

\date{Draft: \today}
\subsection{\Al{} Operation}
\input{supplement_al_figures.tex}

\label{sec:supplement_al}

\AlErrorBudget

The $^{27}\mathrm{Al}^+$ clock operated across nine comparison days under conditions described in \cite{MarshallAlEvaluation2025}, with a single $^{27}\mathrm{Al}^+$ spectroscopy ion co-trapped with a $^{25}\mathrm{Mg}^+$ logic ion for sympathetic cooling and quantum logic readout \cite{SchmidtQLS2005,HumeQLS2007}.  The $^{27}\mathrm{Al}^+$ $^1S_0$$\rightarrow$$^3P_0$ clock transition was interrogated with Rabi spectroscopy and a 1 second pulse time.  Interrogations alternated between Zeeman sublevels with opposite magnetic field dependence to create a magnetic-field independent synthetic average transition, and the probe laser was sent alternately from opposite directions to cancel any first-order Doppler shift.  The fractional systematic uncertainty of this clock was evaluated at $\Delta\nu/\nu=\pm5.5\times10^{-19}$, with leading systematic effects summarized in Table \ref{tab:AlErrorBudget}.  During the measurements reported here, several effects were evaluated and corrected either in real time or after each measurement day, as discussed in \cite{MarshallAlEvaluation2025}; these are starred and their average value across the comparison are given in Table \ref{tab:AlErrorBudget}.  

Between the present measurement campaign and the one reported in \cite{bacon2021}, the $^{27}\mathrm{Al}^+$ clock was completely rebuilt to improve its stability and accuracy.  In light of the revised ratio values presented here, we briefly discuss aspects of the $^{27}\mathrm{Al}^+$ clock and its evaluation that may have differed between the two measurements.  

The new and old $^{27}\mathrm{Al}^+$ clocks used the same optical table and nominal ion-trap position, reducing possible differences in gravitational red shift; measurements with respect to the same USGS survey marker indicate a fractional frequency difference of $\Delta\nu/\nu=(2.7\pm3.3)\times10^{-19}$, consistent with zero change in gravitational red shift between the two apparatus.  

\shiftsbydate

To enable the 1 second probe time, the new clock operates with continuous Doppler cooling applied to the $^{25}\mathrm{Mg}^+$ logic ion, whereas the previous clock was cooled to the motional ground state and probed without additional cooling.  The Doppler-cooled clock has greater motional energy, and therefore a larger second-order Doppler shift, than the ground-state clock.  To evaluate this shift, we made repeated measurements of the motional energy over several months, as described in \cite{MarshallAlEvaluation2025}.  This is the largest systematic uncertainty and second-largest correction in the new clock, so we tested it experimentally during the comparison.  For one comparison day, we adjusted the detuning of our Doppler laser from approximately -20 MHz to -6 MHz to increase the Doppler cooling limit.  We characterized the temperature at the adjusted detuning as being $\sim80~\%$ higher than our usual condition, where we set the detuning of the Doppler laser to the optimal value at half the linewidth of the cooling transition.  The additional fractional frequency shift corresponding to the measured change in secular motional energy was $\Delta\nu/\nu=-8.9\times10^{-18}$, shown in Fig.~ \ref{fig:alshiftsbydate}A.  After applying this correction, both the Al/Yb and Al/Sr ratios agree with the average of the other 8 comparison days, indicating the second-order Doppler shift due to secular motion cannot be a significant source of discrepancy between the previous and current frequency ratio measurements.

The 280 nm Doppler cooling light also applies a Stark shift to the $^{27}\mathrm{Al}^+$ clock transition at 267 nm.  To constrain this shift, we directly measured the polarizability of the clock transition by the 280 nm cooling laser, and calibrated the cooling light intensity by interleaved measurement during clock operation.  

The leading systematic uncertainty in the previous $^{27}\mathrm{Al}^+$ clock \cite{BrewerAlAccuracy2019} was excess micromotion (EMM), fast driven motion at the frequency of the RF trapping potential.  Thanks to an improved trap design, EMM in the new clock is reduced by an order of magnitude or more.  Although both clocks characterized EMM using the resolved-sideband method \cite{KellerMMmeasurement2015}, the evaluation of the EMM shift in the two clocks differed slightly.  In the previous clock, the larger EMM was also less stable, and required active compensation based on measurements on the aluminum ion.  The systematic shift to the clock transition was evaluated by repeatedly running this loop and measuring the resulting EMM amplitude.  In the new clock, the EMM is more stable, and only requires compensation once per day or less.  We therefore measured the EMM before and after each comparison day, taking an average if the two measurements differed and assigning an error bar which includes both measurements as well as zero (see \cite{MarshallAlEvaluation2025} for details).   

Two effects discussed in \cite{bacon2021} were suppressed or eliminated for this new comparison.  First, improved electronic control systems eliminated phase slips in phase-locked loops used for path-length stabilization of light through optical fibers.  In \cite{bacon2021,BrewerAlAccuracy2019} no correction was applied for this effect, but an additional uncertainty of $3\times10^{-19}$ was added based on the clock servo's impulse response to a phase slip.  Second, the average first-order Doppler shift measured in the new apparatus when comparing probes from opposite directions was consistent with zero, at $(0.1\pm 1.7)\times 10^{-18}$; in the previous apparatus, this was approximately $5\times10^{-17}$.  In both cases, averaging the results from the two probe directions suppresses the effect on the measured clock frequency.  However, because of the larger shift, the previous clock was much more sensitive to this cancellation.

The largest systematic shift in both $^{27}\mathrm{Al}^+$ clocks was the dc second-order Zeeman shift due to the quantization magnetic field.  While the magnetic field was 120 $\mu$T in the previous $^{27}\mathrm{Al}^+$ clock and 100 $\mu$T in the new one, the dc Zeeman shift is evaluated in the same way in both clocks---by taking the frequency difference of transitions between Zeeman sublevels of the $^1S_0\rightarrow$$^3P_0$ clock transition with opposite magnetic field dependence.  Over the course of the comparison the second-order Zeeman shift changed by $\Delta\nu/\nu\approx2\times 10^{-17}$, shown in Fig.~\ref{fig:alshiftsbydate}B, without any corresponding change in the corrected ratios. Moreover, previous measurements up to 1 mT agree on this shift at a level below the ratio uncertainty \cite{BrewerMagconst2019}.

%% file: supplement_al_figures.tex
\renewcommand{\figurename}{Fig.}
\renewcommand{\thefigure}{S\arabic{figure}}

\renewcommand{\tablename}{TABLE}
\renewcommand{\thetable}{S\arabic{table}}

\newcommand{\AlErrorBudget}{
\begin{supptable}[htbp!]
\caption{Leading fractional frequency shifts and uncertainties for the NIST $^{27}$Al$^+$ quantum logic clock.  Entries marked * were measured and corrected either in real time or daily; values in this table are the average over the frequency ratio measurement campaign. Shifts in bold are modulated during the comparison.  Gravitational redshift is measured to NGS marker 1H116 \cite{NGSSurvey2019}.  See \cite{MarshallAlEvaluation2025} for more details.}
\begin{tabular}{lcc}
\hline \hline
Effect & Shift ($10^{-19}$) & Uncertainty ($10^{-19}$) \\ \hline
  \textbf{Secular motion}     &           -114.6         &           3.8         \\ 
  dc quad. Zeeman* & -6317.9 & 2.5 \\
  Cooling laser Stark* & -37.2 & 2.0 \\
  Blackbody radiation & -30.7 & 1.7 \\
  Excess micromotion* & -1.6 & 1.6 \\
  All other effects & -0.8 & 0.8 \\
  \hline
  Total & -6502.8 & 5.5 \\ \hline \hline
  Redshift to marker & 1504.9 & 3.3\\
  Total with redshift & -4997.9 & 6.4\\ \hline \hline
\label{tab:AlErrorBudget}
\end{tabular}
\end{supptable}
}

\newcommand{\shiftsbydate}{
\begin{suppfigure}[tbp!]
\begin{center}
\includegraphics*[width=\columnwidth]{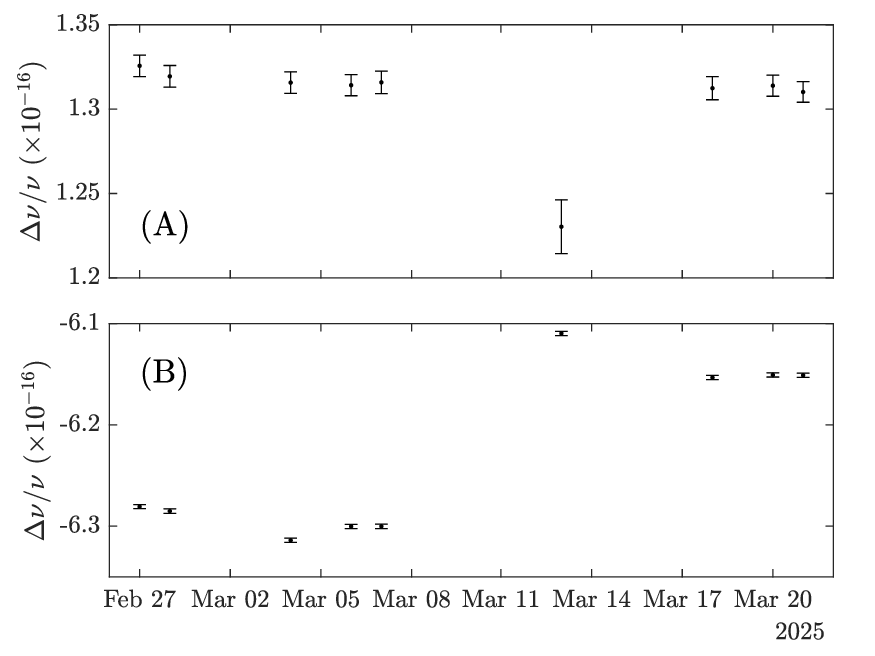}
\caption{Systematic shifts applied to each day of $^{27}\mathrm{Al}^+$ data.  \textbf{(A)} All shifts except dc quadratic Zeeman effect, including gravitational redshift between the ion trap position and the nearby USGS geopotential survey marker.  The large change on Mar.~13 reflects the increased Doppler temperature.  \textbf{(B)} Second-order Zeeman shift due to the quantization field.  Vertical scale corresponds to a magnetic field range of approximately 1.5 $\mu$T.}
\label{fig:alshiftsbydate}
\end{center}
\end{suppfigure}
}

%% file: Yb_Supplement.tex
% \documentclass[%
%  reprint,
% superscriptaddress,
% %groupedaddress,
% %unsortedaddress,
% %runinaddress,
% %frontmatterverbose, 
% %preprint,
% %showpacs,preprintnumbers,
% %nofootinbib,
% %nobibnotes,
% %bibnotes,
%  amsmath,amssymb,
%  aps,
% %pra,
% %prb,
% %rmp,
% %prstab,
% %prstper,
% %floatfix,
% float]{revtex4-1}

% \usepackage{subfiles}

% \usepackage{graphicx}% Include figure files
% \usepackage{dcolumn}% Align table columns on decimal point
% \usepackage{bm}% bold math
% \usepackage{amsmath}
% \usepackage{physics}
% \usepackage{gensymb}
% \usepackage{placeins}
% %\usepackage{hyperref}% add hypertext capabilities
% %\usepackage[mathlines]{lineno}% Enable numbering of text and display math
% %\linenumbers\relax % Commence numbering lines

% %\usepackage[showframe,%Uncomment any one of the following lines to test 
% %%scale=0.7, marginratio={1:1, 2:3}, ignoreall,% default settings
% %%text={7in,10in},centering,
% %%margin=1.5in,
% %%total={6.5in,8.75in}, top=1.2in, left=0.9in, includefoot,
% %%height=10in,a5paper,hmargin={3cm,0.8in},
% %]{geometry}

% % Redefine \maketitle so that it can be used twice (for supplementary)
% \makeatletter
% \def\maketitle{
% \@author@finish
% \title@column\titleblock@produce
% \suppressfloats[t]}
% \makeatother

% \usepackage[colorlinks=true, linkcolor=blue, urlcolor=blue, citecolor=blue]{
% 	hyperref
% }% add hypertext capabilities
% \usepackage[mathlines]{lineno}% Enable numbering of text and display math
% \linenumbers\relax % Commence numbering lines
% \begin{document}
	%%%%%%%%%%  Custom commands %%%%%%%%%%
	% \newcommand{\Yb}{${}^{171}$Yb}
	\newcommand{\Er}{$E_{r}$}
	\renewcommand{\thefigure}{S\arabic{figure}}
	\makeatother
	\renewcommand{\thetable}{S\arabic{table}}
	\makeatother

	%%%%%%%%%%%%%%%%%%%%%%%%%%%%%%%%%%%%%%

	% \preprint{}
	\title{Supplementary Material: BACON 2025 Yb Notes, still very much a work in progress by Jacob, don't take very seriously yet}
	% \maketitle
    
\subsection{\Yb{} OLC Operation}
\emph{\textbf{Operational conditions.}} For this work, the Yb clock was operated similar to the BACON21 experiment, detailed in Refs.~\cite{bacon2021, mcgrew_Yb2018}.
Much of the hardware and software from BACON21 was unchanged, as were 
systematic shift evaluations from servo error, frequency synthesis, tunneling, line pulling, probe Stark, residual Doppler, and DC Stark effects.
Most operating parameters were also similar to BACON21, such as the use of a  50\Er ~lattice depth, a total atom number of approximately 1000, and use of a room-temperature BBR environment.
However, as described below, we use more recent evaluations of systematic shifts from lattice Stark, atomic density, and BBR Stark.
Additionally, we reevaluate the height of the atoms, leading to a slight difference in gravitational redshift from BACON21. We also improved the vacuum lifetime, which reduces the background gas collision shift.  
The total uncertainty of the Yb clock for typical operating conditions is $2.6\times10^{-18}$, dominated by lattice light shift uncertainty as shown in Tab.~\ref{tab:YbErrorBudget}.

We note that the newly updated systematic analysis of BBR and lattice light shifts, discussed below, modestly affect the agreement with BACON21.
These updated systematic analyses are based on new determinations of atomic coefficients that represent inherent properties of the Yb atom. 
Thus, they are expected to remain constant over the time between this work and BACON21, the calculated statistical agreement between this work and BACON21 should use a consistent set of atomic coefficients (the updated ones) to determine both ratios. 
As such, we rework the BACON21 analysis using the updated systematic effects from this work.
Only the Yb frequency is updated as there are no changes to any Al\textsuperscript{+} atomic coefficients between BACON21 and this work. 
This changes the Yb frequency in BACON21 by $-1.5\times10^{-18}$.
This is negligible relative to both the observed change in ratios from BACON21 as well as the uncertainties in the BACON21 ratio values.
The statistical agreement of the Al\textsuperscript{+}/Yb ratio between this work and the ratio as written directly in BACON21 is 2.4$\sigma$.
When using the updated systematic effects to rework the BACON21 ratio the statistical agreement is 2.6$\sigma$.

\begin{supptable}[htbp!]
\caption{Leading fractional frequency shifts and uncertainties for the NIST $^{171}$Yb clock. Values marked with * are corrected for in postprocessing based on real-time measurements. The reported values below are for typical conditions. Shifts in bold are modulated during the comparison. The gravitational redshift is relative to NGS survey marker 811G104. }
\begin{tabular}{lcc}
\hline \hline
Effect & Shift ($10^{-18}$) & Uncertainty ($10^{-18}$) \\ \hline

  Lattice light     &           -1.3         &           2.4        \\ 
  BBR* & -2361.2 & 0.7 \\
  Background gas & -3.1 & 0.3 \\
  Density* & -0.9 & 0.3  \\
  
  \textbf{2nd order Zeeman}* & -118.6 & 0.2 \\
  All other effects & 0.01 & 0.2 \\ \hline
  Total & 2485.1 & 2.6\\
 \hline \hline
  Gravitational redshift & 152.6 & 0.5 \\
  Total with redshift & 2332.5 & 2.6 \\ \hline \hline
\label{tab:YbErrorBudget}
\end{tabular}
\end{supptable}

\emph{\textbf{New Lattice Light Shift Analysis.}} For the measurements reported here, we perform a quasi-independent evaluation of the lattice light shift.
For comparison, the original BACON21 lattice light shift evaluation and operation, detailed in \cite{mcgrew_Yb2018}, used longitudinal sideband cooling, relatively warmer radial temperatures, and a thermally-averaged model of the lattice light shift.  That work also accounted for magnetic dipole and electric quadrupole polarizability ($\tilde{\alpha}^{\mathrm{M1E2}}$) effects based on atomic structure theory calculations.
However, as noted in BACON21, an experimental measurement of $\tilde{\alpha}^{\mathrm{M1E2}}$ suggested a much larger effect \cite{nemitz_2019}, which would correspond to an additional light shift of $3 \times 10^{-18}$.  Since that time, errors in the atomic structure calculations from coupling to negative energy states have been resolved. We have also carried out an experimental measurement of $\tilde{\alpha}^{\mathrm{M1E2}}$ \cite{Bothwell_2025}, nominally in agreement with that of \cite{nemitz_2019}, though with both a reduced mean value and smaller uncertainty.

The lattice light shift evaluation here benefits from our recent work \cite{Bothwell_2025}, using the improved values of $\tilde{\alpha}^{\mathrm{M1E2}}$ and $\frac{\partial\tilde{\alpha}^{\mathrm{E1}}}{\partial\nu}$ reported therein. However, here we add new experimental measurements of the lattice light shift, in order to improve the determinations of hyperpolarizability, $\tilde{\beta}$, and the magic frequency, $\nu_{\mathrm{E1}}$, compared to that work.  Like our previous evaluation from Ref.~\cite{mcgrew_Yb2018}, we employ longitudinal sideband cooling.
However, this time we also employ Sisyphus cooling \cite{Chen24} to lower the radial temperature to $\sim450$ nK.  We also combined data taken on two different Yb lattice clocks, YbI and YbII.
To measure the light shifts, we employed the traditional method of self-interleaving different trap depths.  The self-interleaved measurement stability generally benefited from the low-noise silicon cavity, as good as $2.5\times10^{-16}/ \sqrt{\tau}$, with $\tau$ the averaging time in seconds.    

As before, we evaluate density shifts at various lattice depths to correct for this potential systematic bias. 
For the deepest depths investigated here ($>400$\Er{}), the density shifts are relatively large and generally over-scattered, yielding a lattice light shift uncertainty that exceeded the statistical uncertainty of most lattice light shift measurements we made. 
Below 400\Er{} the density shift uncertainty is at or below the statistical uncertainty of a single measurement, and in this regime we applied a $U_0^{5/4}$ density shift scaling to separate lower trap-depth data.  
Typical atom numbers for these measurements were 1000 for YbII and 500 for YbI (note that YbI has a tighter lattice waist).

The self-interleaved measurements, with density shift corrections, are plotted in Figure \ref{fig:YbNewLatticeLight}.
The lower of the two trap depths was typically at a value between 20\Er{} and 35\Er{}, with the higher depth reaching as high as 440\Er{}. 
The difference in interleaved trap depths, $\Delta U_0$, is plotted on the horizontal axis.  We use a Monte-Carlo method to simultaneously fit the YbI and YbII data with the model of \cite{ushijima_ls_2018}, and include $\frac{\partial\tilde{\alpha}^{\mathrm{E1}}}{\partial\nu}$ and $\tilde{\alpha}^{\mathrm{M1E2}}$ from Ref. \cite{Bothwell_2025}.
We find $\nu_{\mathrm{E1}}=394,798,263.8(1.5)$ MHz and $\tilde{\beta}=-1.2(2)\times10^{-21}$.
We have inflated the error bars by $\sqrt{\chi_{red}^2}=1.75$. 
We note that $\nu_{\mathrm{E1}}$ is in $1.3\sigma$ agreement with that obtained by fitting the YbII data only, and is in $1\sigma$ agreement with the values we measured independently in \cite{Bothwell_2025} and \cite{mcgrew_Yb2018}.

Based on the $\nu_{\mathrm{E1}}$ and $\tilde{\beta}$ measured here and $\tilde{\alpha}^{\mathrm{M1E2}}$ and $\frac{\partial\tilde{\alpha}^{\mathrm{E1}}}{\partial\nu}$ measured in Ref.~\cite{Bothwell_2025}, we find a lattice light shift of $-1.3(2.4)\times10^{-18}$ at our typical operating conditions of $U_0=50(2)$$E_r$, $T_r=1.1(2)$ $\mu$K, $n_z=0.05(2)$, and $\delta_L\approx5.1$ MHz.  For comparison, the BACON21 evaluation would yield a lattice light shift of $-3.2(1.1)\times10^{-18}$ for these same conditions, consistent below the 1$\sigma$ level.

\begin{suppfigure}
    \centering
    \includegraphics[width=0.99\linewidth]{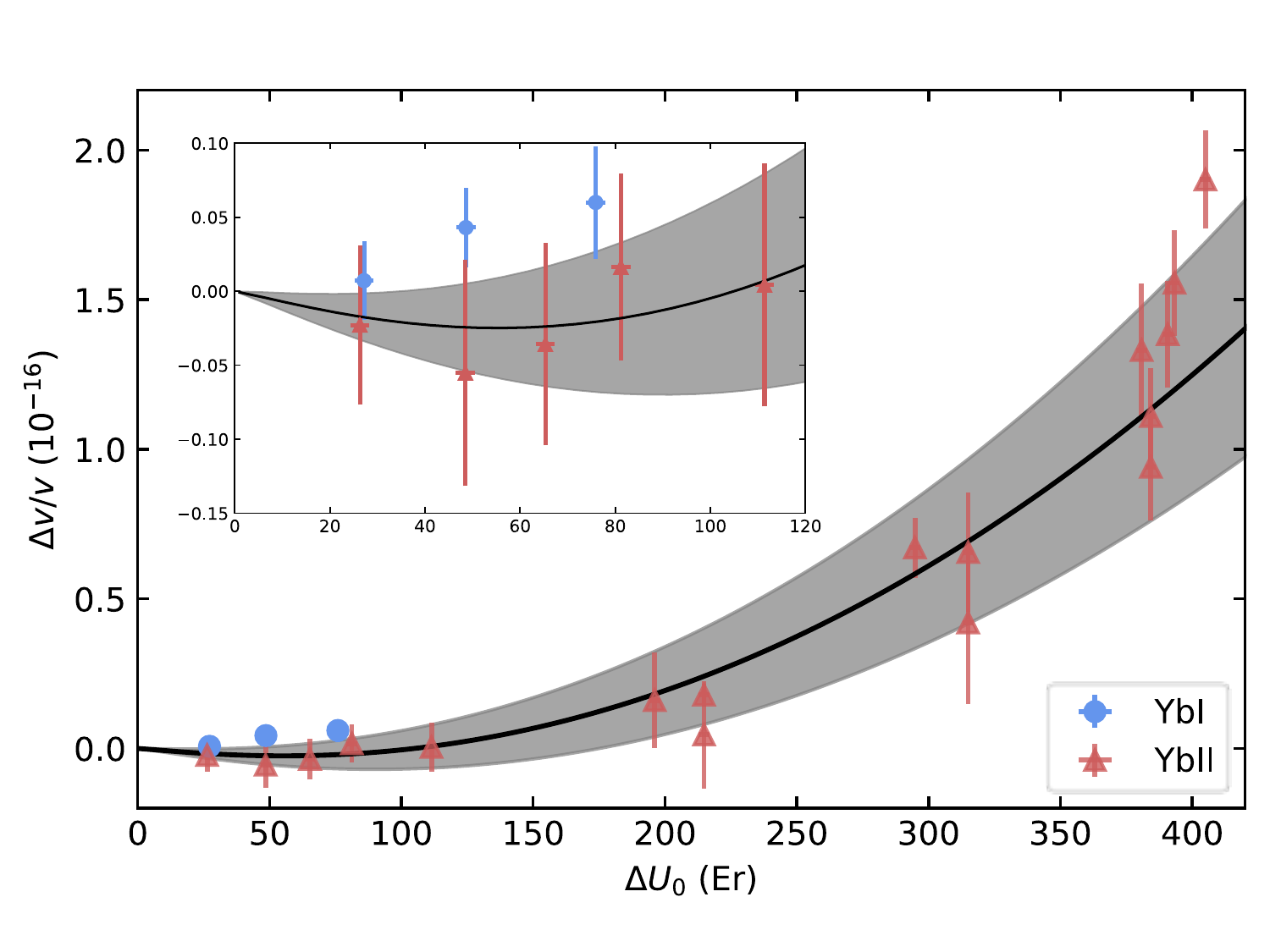}
    \caption{The lattice light shift parameters $\nu_{\mathrm{E1}}$ and $\tilde{\beta}$ are measured. 
    The measured shift, including density shift corrections, is plotted versus the difference in trap depths we interleaved between. 
    The fitted lattice light shift, with shaded 1$\sigma$ confidence interval is plotted in black. The confidence interval includes the statistical uncertainty of the fit and the uncertainties of $\frac{\partial\tilde{\alpha}^{\mathrm{E1}}}{\partial\nu}$ and $\tilde{\alpha}^{\mathrm{M1E2}}$, which are constrained to the values from Ref.~\cite{Bothwell_2025}.
    The inset is a magnified scale of the low depth data.}
    \label{fig:YbNewLatticeLight}
\end{suppfigure}

\emph{\textbf{Other systematic effects.}} Due to slow improvement in the system vacuum pressure, the vacuum lifetime has been re-measured at $5.3(1)$ s.  This corresponds to a background gas collision shift of $-3.1(2)\times10^{-18}$, which is both smaller in magnitude and uncertainty than in Ref. \cite{mcgrew_Yb2018}. 

The gravitational redshift of the  \textsuperscript{171}Yb atoms relative to the \textsuperscript{27}Al\textsuperscript{+} and \textsuperscript{87}Sr clocks is made via height measurements of the Yb atoms relative to a survey marker and the reported relative heights between markers~\cite{NGSSurvey2019}.  
The resulting redshift uncertainty is $5\times10^{-19}$.

The dynamic contribution of the BBR shift uncertainty has recently been improved (30~\% reduction) by taking the weighted mean of our new experimentally measured value with previous theoretical and semi-empirical values \cite{hassan_2025}.
We note that the new dynamic BBR shift is in 0.5$\sigma$ agreement with the value used in BACON21.
For temperatures of the clock on a typical day, the BBR shift uncertainty is improved from $9\times10^{-19}$ in BACON21 to $7\times10^{-19}$ in this work.

The atomic density shift depends on atom number, lattice confinement, atomic temperature, and atomic distribution within the lattice.  Without a careful assessment of how each of these factors change over time, we have re-measured a density shift of $-0.88(30)\times10^{-18}$ per 1000 atoms at 50 $E_r$,  as detailed in Ref.~\cite{Siegel2025}. 
Including a 10~\% uncertainty in atom number, a typical density shift uncertainty is then $3\times10^{-19}$.
This uncertainty is similar to Ref.~\cite{mcgrew_Yb2018} when including both $s$- and $p$-wave shift effects, though we note that the magnitude of the resulting shift in that work was smaller, only $-0.21\times10^{-18}$.

\emph{\textbf{Systematics modulation.}} To experimentally probe the validity of some systematic shifts on the Yb clock, the magnetic bias field and the clock cycle time were adjusted. 
A typical magnetic bias field of 1 G is applied on all nominal days, splitting the magnetic sub-levels by $\sim400$ Hz.
On March 18th, we instead operated the system with a 2 G bias field, as shown in Fig.~\ref{fig:YbSystematics}.
Frequency comparisons to \textsuperscript{87}Sr and \textsuperscript{27}Al\textsuperscript{+} revealed no visible change in the corrected Yb clock frequency, validating the applied second order Zeeman shift correction and cancellation of the first order Zeeman shift (through interrogation of alternating Zeeman states) at the low $10^{-18}$ level.
On March 18th, we also added an additional 200 ms of extra lattice holding time prior to spectroscopy.  This can be one way to investigate possible first order Doppler shifts or optical-local-oscillator-induced biases from clock-cycle synchronous effects.  
No change in the measured optical ratios was observed at the low $10^{-18}$ level.

\begin{suppfigure}
    \centering
    \includegraphics[width=0.99\linewidth]{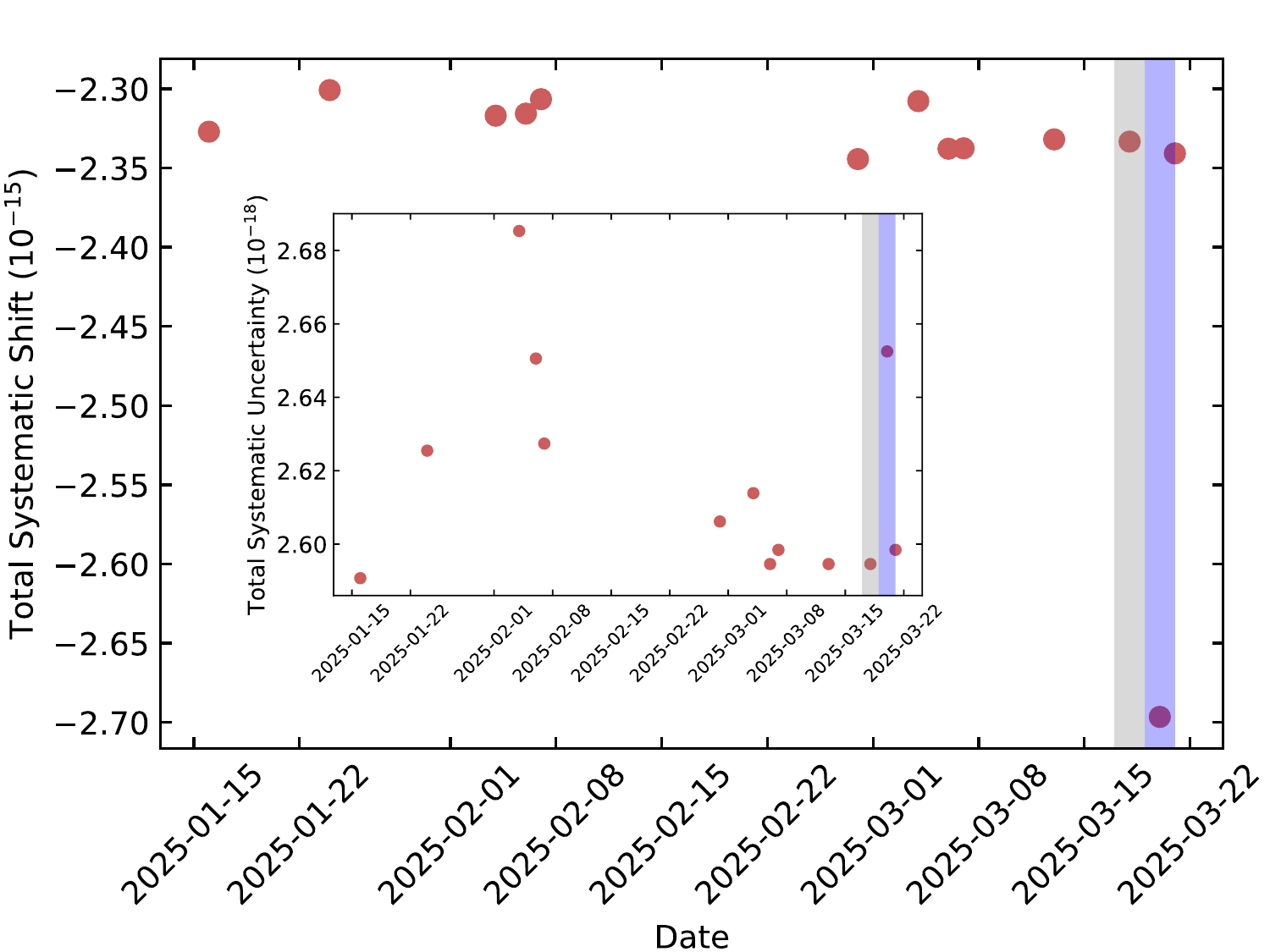}
    \caption{The total systematic shift of the Yb clock including the gravitational redshift to the local geodetic reference marker.
    March 20th, twice nominal magnetic bias field, is shaded blue and March 18th, extra 200 ms cycle time, is shaded gray.  
    The inset shows how the total systematic uncertainty varies over the measurement campaign. 
    Most of the variation, especially before February 22nd, is due to varying atom numbers affecting the density shift. 
    }
    \label{fig:YbSystematics}
\end{suppfigure}

\emph{\textbf{Real-time BBR shift corrections.}} The Yb clock is furnished with an in-vacuum copper BBR ``shield,'' enclosing the atoms in a near-ideal BBR environment \cite{beloy_2014}. 
The shield is not actively temperature controlled; however, seven thermometers monitor the temperature of the shield in real time. 
These real-time temperature measurements, along with geometric knowledge about the shield, allow for real-time corrections of the BBR shift. 
For high accuracy operation of the Yb clock, as was done in this work, we typically run the system the night before a ratio measurement, to reduce `warm-up' thermal drifts of the shield during measurements.
Doing so gives shield temperatures that are generally stable at the $\sim100$ mK level.

On January 24th, the warm-up procedure is omitted. 
As a result, the temperature of the shield drifted by several Kelvin over the $\sim4$ hour ratio measurement. 
While we estimate that this adds a negligible $3\times10^{-19}$ uncertainty to the Yb clock \cite{Siegel2025}, it gave us an opportunity to test the fidelity of our real-time BBR corrections using the \YbSr{} ratio.
Fig.~\ref{fig:Yb_BBR} plots this ratio with and without the real-time BBR corrections applied.  Without them, the ratio exhibits a sizable drift of $\sim6\times10^{-17}$, virtually identical to the real-time BBR corrections. 
We note that, using multiple Yb clocks, we have verified the BBR shift correction over a wide temperature range varying between $\sim70$ K and $\sim320$ K on YbII at $<2\times10^{-18}$ \cite{hassan_2025, Siegel2025}.

\begin{suppfigure}
    \centering
    \includegraphics[width=0.99\linewidth]{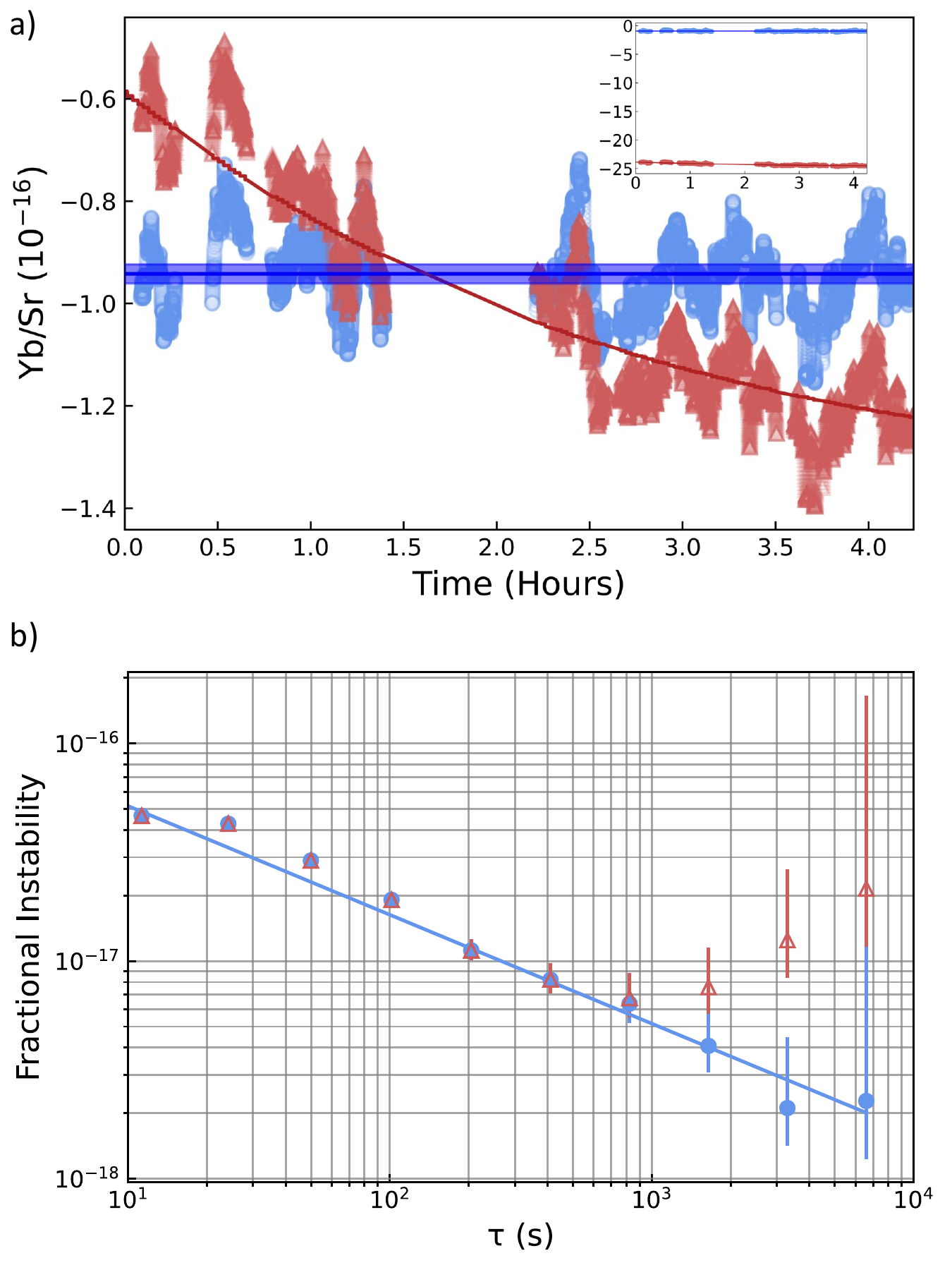}
    \caption{The real-time Yb BBR shift correction is validated using the Yb/Sr ratio on January 24th, 2025. 
    a) We plot the \Yb/\textsuperscript{87}Sr ratio as a fractional offset from the 2021 CIPM recommended ratio. 
    The corrections with a 311 s rolling average are plotted as blue circles. 
    Red triangles show the ratio if only the average temperature over the run is used to make a BBR correction for all points.
    The blue line shows the average ratio for the day, with shaded regions as a 1-$\sigma$ confidence interval.
    The red line shows the real-time temperature corrections made, which refine the average-over-the-whole-run temperature correction.
    The inset shows blue points as the real-time corrected ratio, and red points as the ratio without any temperature corrections, corresponding to the magnitude of the room-temperature BBR shift in Yb.
    b) The fractional instability of the \Yb/\textsuperscript{87}Sr ratio is plotted in blue circles for real-time BBR corrections and in red triangles for the average-over-the-whole-run BBR correction. 
    The blue line is a $1/\sqrt{\tau}$ fit to the real-time corrected data after 100 s.}
    \label{fig:Yb_BBR}
\end{suppfigure}

%% file: main.bbl
\begin{thebibliography}{10}

\bibitem{Weyers_2018}
S. Weyers, V. Gerginov, M. Kazda, J. Rahm, B. Lipphardt, G. Dobrev, and K. Gibble, Metrologia {\bf 55},  789  (2018).

\bibitem{GerginovNISTF42025}
V. Gerginov, G.~W. Hoth, T.~P. Heavner, T.~E. Parker, K. Gibble, and J.~A. Sherman, Metrologia {\bf 62},  035002  (2025).

\bibitem{BeattieCanadaCs2025}
S. Beattie, B. Jian, C. Marceau, K. Gibble, and M. Gertsvolf, Metrologia {\bf 62},  035003  (2025).

\bibitem{LudlowOpticalClockReview2015}
A.~D. Ludlow, M.~M. Boyd, J. Ye, E. Peik, and P. Schmidt, Reviews of Modern Physics {\bf 87},  637–701  (2015).

\bibitem{Dimarcq_2024}
N. Dimarcq, M. Gertsvolf, G. Mileti, S. Bize, C.~W. Oates, E. Peik, D. Calonico, T. Ido, P. Tavella, F. Meynadier, G. Petit, G. Panfilo, J. Bartholomew, P. Defraigne, E.~A. Donley, P.~O. Hedekvist, I. Sesia, M. Wouters, P. Dub{\'e}, F. Fang, F. Levi, J. Lodewyck, H.~S. Margolis, D. Newell, S. Slyusarev, S. Weyers, J.-P. Uzan, M. Yasuda, D.-H. Yu, C. Rieck, H. Schnatz, Y. Hanado, M. Fujieda, P.-E. Pottie, J. Hanssen, A. Malimon, and N. Ashby, Metrologia {\bf 61},  012001  (2024).

\bibitem{schioppo_comparing_2022}
M. Schioppo, J. Kronjäger, A. Silva, R. Ilieva, J.~W. Paterson, C.~F.~A. Baynham, W. Bowden, I.~R. Hill, R. Hobson, A. Vianello, M. Dovale-Álvarez, R.~A. Williams, G. Marra, H.~S. Margolis, A. Amy-Klein, O. Lopez, E. Cantin, H. Álvarez Martínez, R. Le~Targat, P.~E. Pottie, N. Quintin, T. Legero, S. Häfner, U. Sterr, R. Schwarz, S. Dörscher, C. Lisdat, S. Koke, A. Kuhl, T. Waterholter, E. Benkler, and G. Grosche, Nature Communications {\bf 13},  212  (2022).

\bibitem{grebingRealizationTimescaleAccurate2016}
C. Grebing, A. {Al-Masoudi}, S. D{\"o}rscher, S. H{\"a}fner, V. Gerginov, S. Weyers, B. Lipphardt, F. Riehle, U. Sterr, and C. Lisdat, Optica {\bf 3},  563  (2016).

\bibitem{hachisuMonthslongRealtimeGeneration2018}
H. Hachisu, F. Nakagawa, Y. Hanado, and T. Ido, Scientific Reports {\bf 8},  4243  (2018).

\bibitem{yaoOpticalClockBasedTimeScale2019}
J. Yao, J.~A. Sherman, T. Fortier, H. Leopardi, T. Parker, W. McGrew, X. Zhang, D. Nicolodi, R. Fasano, S. Sch{\"a}ffer, K. Beloy, J. Savory, S. Romisch, C. Oates, S. Diddams, A. Ludlow, and J. Levine, Physical Review Applied {\bf 12},  044069  (2019).

\bibitem{milnerDemonstrationTimescaleBased2019}
W.~R. Milner, J.~M. Robinson, C.~J. Kennedy, T. Bothwell, D. Kedar, D.~G. Matei, T. Legero, U. Sterr, F. Riehle, H. Leopardi, T.~M. Fortier, J.~A. Sherman, J. Levine, J. Yao, J. Ye, and E. Oelker, Physical Review Letters {\bf 123},  173201  (2019).

\bibitem{bacon2021}
{Boulder Atomic Clock Optical Network {(BACON)} Collaboration}, Nature {\bf 591},  564  (2021).

\bibitem{HausserIndiumMultiIon2025}
H.~N. Hausser, J. Keller, T. Nordmann, N.~M. Bhatt, J. Kiethe, H. Liu, I.~M. Richter, M. Von~Boehn, J. Rahm, S. Weyers, E. Benkler, B. Lipphardt, S. Dörscher, K. Stahl, J. Klose, C. Lisdat, M. Filzinger, N. Huntemann, E. Peik, and T.~E. Mehlstäubler, Physical Review Letters {\bf 134},  023201  (2025).

\bibitem{SafronovaNewPhysicsReview2018}
M.~S. Safronova, D. Budker, D. DeMille, D.~F.~J. Kimball, A. Derevianko, and C.~W. Clark, Reviews of Modern Physics {\bf 90},  025008  (2018).

\bibitem{KennedyDarkMatterClock2020}
C.~J. Kennedy, E. Oelker, J.~M. Robinson, T. Bothwell, D. Kedar, W.~R. Milner, G.~E. Marti, A. Derevianko, and J. Ye, Physical Review Letters {\bf 125},  201302  (2020).

\bibitem{FilzingerClocksDM2023}
M. Filzinger, S. Dörscher, R. Lange, J. Klose, M. Steinel, E. Benkler, E. Peik, C. Lisdat, and N. Huntemann, Physical Review Letters {\bf 130},  253001  (2023).

\bibitem{LangePositionInvariance2021}
R. Lange, N. Huntemann, J.~M. Rahm, C. Sanner, H. Shao, B. Lipphardt, C. Tamm, S. Weyers, and E. Peik, Physical Review Letters {\bf 126},  011102  (2021).

\bibitem{SafronovaFundamentalConstantVariation2019}
M.~S. Safronova, Annalen der Physik {\bf 531},  1800364  (2019).

\bibitem{SherrillFundamentalConstantVariation2023}
N. Sherrill, A.~O. Parsons, C.~F.~A. Baynham, W. Bowden, E. Anne~Curtis, R. Hendricks, I.~R. Hill, R. Hobson, H.~S. Margolis, B.~I. Robertson, M. Schioppo, K. Szymaniec, A. Tofful, J. Tunesi, R.~M. Godun, and X. Calmet, New Journal of Physics {\bf 25},  093012  (2023).

\bibitem{Flury_relGeo_2016}
J. Flury, Journal of Physics: Conference Series {\bf 723},  012051  (2016).

\bibitem{grotti_chronometric_2024}
J. Grotti, I. Nosske, S. Koller, S. Herbers, H. Denker, L. Timmen, G. Vishnyakova, G. Grosche, T. Waterholter, A. Kuhl, S. Koke, E. Benkler, M. Giunta, L. Maisenbacher, A. Matveev, S. D\"orscher, R. Schwarz, A. Al-Masoudi, T. H\"ansch, T. Udem, R. Holzwarth, and C. Lisdat, Phys. Rev. Appl. {\bf 21},  L061001  (2024).

\bibitem{MarshallAlEvaluation2025}
M.~C. Marshall, D.~A.~R. Castillo, W.~J. Arthur-Dworschack, A. Aeppli, K. Kim, D. Lee, W. Warfield, J. Hinrichs, N.~V. Nardelli, T.~M. Fortier, J. Ye, D.~R. Leibrandt, and D.~B. Hume, Phys. Rev. Lett. {\bf 135},  033201  (2025).

\bibitem{mcgrew_Yb2018}
W.~F. McGrew, X. Zhang, R.~J. Fasano, S.~A. Sch{\"a}ffer, K. Beloy, D. Nicolodi, R.~C. Brown, N. Hinkley, G. Milani, M. Schioppo, T.~H. Yoon, and A.~D. Ludlow, Nature {\bf 564},  87  (2018).

\bibitem{aeppli_SrAccuracy}
A. Aeppli, K. Kim, W. Warfield, M.~S. Safronova, and J. Ye, Phys. Rev. Lett. {\bf 133},  023401  (2024).

\bibitem{SeeSupplementalMaterial}
See {{Supplemental Material}} at [{{URL}} Will Be Inserted by Publisher] for More Information.

\bibitem{mateiMmLasersSub102017}
D.~G. Matei, T. Legero, S. H{\"a}fner, C. Grebing, R. Weyrich, W. Zhang, L. Sonderhouse, J.~M. Robinson, J. Ye, F. Riehle, and U. Sterr, Physical Review Letters {\bf 118},  263202  (2017).

\bibitem{oelker_Si2019}
E. Oelker, R.~B. Hutson, C.~J. Kennedy, L. Sonderhouse, T. Bothwell, A. Goban, D. Kedar, C. Sanner, J.~M. Robinson, G.~E. Marti, D.~G. Matei, T. Legero, M. Giunta, R. Holzwarth, F. Riehle, U. Sterr, and J. Ye, Nature Photonics {\bf 13},  714  (2019).

\bibitem{BressiNeutrality2011}
G. Bressi, G. Carugno, F. Della~Valle, G. Galeazzi, G. Ruoso, and G. Sartori, Physical Review A {\bf 83},  052101  (2011).

\bibitem{bran_documentation}
J. Van~Dyke, BRAN Technical Information, 2004.

\bibitem{Ye_BRAN_03}
J. Ye, J.-L. Peng, R.~J. Jones, K.~W. Holman, J.~L. Hall, D.~J. Jones, S.~A. Diddams, J. Kitching, S. Bize, J.~C. Bergquist, L.~W. Hollberg, L. Robertsson, and L.-S. Ma, J. Opt. Soc. Am. B {\bf 20},  1459  (2003).

\bibitem{narbonneauHighResolutionFrequency2006}
F. Narbonneau, M. Lours, S. Bize, A. Clairon, G. Santarelli, O. Lopez, {\relax Ch}. Daussy, A. {Amy-Klein}, and {\relax Ch}. Chardonnet, Review of Scientific Instruments {\bf 77},  064701  (2006).

\bibitem{Nardelli_2023}
N.~V. Nardelli, H. Leopardi, T.~R. Schibli, and T.~M. Fortier, Laser \& Photonics Reviews {\bf 17},  2200650  (2023).

\bibitem{FortierTiSaph2006}
T.~M. Fortier, A. Bartels, and S.~A. Diddams, Optics Letters {\bf 31},  1011–1013  (2006).

\bibitem{FortierCombReview2019}
T. Fortier and E. Baumann, Communications Physics {\bf 2},  153  (2019).

\bibitem{NemitzYbSrRatio2016}
N. Nemitz, T. Ohkubo, M. Takamoto, I. Ushijima, M. Das, N. Ohmae, and H. Katori, Nature Photonics {\bf 10},  258–261  (2016).

\bibitem{LindvallClockComparisons2025}
T. Lindvall, M. Pizzocaro, R. Godun, M. Abgrall, D. Akamatsu, A. Amy-Klein, E. Benkler, N. Bhatt, D. Calonico, E. Cantin, E. Cantoni, G. Cerretto, C. Chardonnet, M.~A.~C. Marin, C. Clivati, S. Condio, E.~A. Curtis, H. Denker, S. Donadello, S. Dörscher, C.-H. Feng, M. Filzinger, T. Fordell, I. Goti, K. Hanhijärvi, H.~N. Hausser, I. Hill, K. Hosaka, N. Huntemann, M. Johnson, J. Keller, J. Klose, T. Kobayashi, S. Koke, A. Kuhl, R.~L. Targat, T. Legero, F. Levi, B. Lipphardt, C. Lisdat, H. Liu, J. Lodewyck, O. Lopez, M. Mazouth-Laurol, T. Mehlstaeubler, A. Mura, A. Nishiyama, T. Nordmann, A. Parsons, G. Petit, B. Pointard, P.-E. Pottie, M. Risaro, B. Robertson, M. Schioppo, H. Shang, K. Stahl, M. Steinel, U. Sterr, A. Tofful, M. Tønnes, D.~B.~A. Tran, J. Tunesi, A. Wallin, and H. Margolis, Optica {\bf 12},  843  (2025).

\bibitem{margolisCIPMListRecommended2024}
H.~S. Margolis, G. Panfilo, G. Petit, C. Oates, T. Ido, and S. Bize, Metrologia {\bf 61},  035005  (2024).

\bibitem{bothwellJILASrIOptical2019}
T. Bothwell, D. Kedar, E. Oelker, J.~M. Robinson, S.~L. Bromley, W.~L. Tew, J. Ye, and C.~J. Kennedy, Metrologia {\bf 56},  065004  (2019).

\bibitem{aeppliHamiltonianEngineeringSpinorbitcoupled2022}
A. Aeppli, A. Chu, T. Bothwell, C.~J. Kennedy, D. Kedar, P. He, A.~M. Rey, and J. Ye, Science Advances {\bf 8},  eadc9242  (2022).

\bibitem{BrewerAlAccuracy2019}
S.~M. Brewer, J.-S. Chen, A.~M. Hankin, E.~R. Clements, C.-W. Chou, D.~J. Wineland, D.~B. Hume, and D.~R. Leibrandt, Physical Review Letters {\bf 123},  033201  (2019).

\bibitem{BrewerMagconst2019}
S.~M. Brewer, J.-S. Chen, K. Beloy, A.~M. Hankin, E.~R. Clements, C.~W. Chou, W.~F. McGrew, X. Zhang, R.~J. Fasano, D. Nicolodi, H. Leopardi, T.~M. Fortier, S.~A. Diddams, A.~D. Ludlow, D.~J. Wineland, D.~R. Leibrandt, and D.~B. Hume, Physical Review A {\bf 100},  013409  (2019).

\bibitem{Siegel2025}
J.~L. Siegel, in preparation  (2025).

\bibitem{KoepkeConsensusBuilder2017}
A. Koepke, T. Lafarge, A. Possolo, and B. Toman, Metrologia {\bf 54},  S34  (2017).

\bibitem{ThompsonEllisonDarkUncertainty2011}
M. Thompson and S.~L.~R. Ellison, Accreditation and Quality Assurance {\bf 16},  483–487  (2011).

\bibitem{BodnarHeterogeneity2020}
O. Bodnar, R. Nalule~Muhumuza, and A. Possolo, Metrologia {\bf 57},  064004  (2020).

\bibitem{Note1}
Notably, the JILA-NIST link was tested differently in the two campaigns: in BACON21{}, a loopback test was performed only at $1.5~\mu $m, including the 3.6-km link and NIST frequency combs but omitting elements solely at JILA, whereas in the current campaign, the test extends to 698~nm as shown in Fig.~\ref {fig:fig3}(b).

\bibitem{GrayTCH1974}
J. Gray and D. Allan,  in {\em 28th {{Annual Symposium}} on {{Frequency Control}}} (IEEE, Atlantic City, NJ, USA, 1974), pp.\ 243--246.

\bibitem{Note2}
See for example the missing points from Fig.~\ref {fig:tchfig}(b) around $10^4$ s, where the model calculates an unphysical negative variance \cite {GrayTCH1974} and only an upper bound can be plotted.

\bibitem{KimDiffspec2023}
M.~E. Kim, W.~F. McGrew, N.~V. Nardelli, E.~R. Clements, Y.~S. Hassan, X. Zhang, J.~L. Valencia, H. Leopardi, D.~B. Hume, T.~M. Fortier, A.~D. Ludlow, and D.~R. Leibrandt, Nature Physics {\bf 19},  25  (2023).

\bibitem{akamatsuFrequencyRatioMeasurement2014}
D. Akamatsu, M. Yasuda, H. Inaba, K. Hosaka, T. Tanabe, A. Onae, and F.-L. Hong, Optics Express {\bf 22},  7898  (2014).

\bibitem{campbellAbsoluteFrequencyThe87Sr2008}
G.~K. Campbell, A.~D. Ludlow, S. Blatt, J.~W. Thomsen, M.~J. Martin, M.~H.~G. {de Miranda}, T. Zelevinsky, M.~M. Boyd, J. Ye, S.~A. Diddams, T.~P. Heavner, T.~E. Parker, and S.~R. Jefferts, Metrologia {\bf 45},  539  (2008).

\bibitem{grottiGeodesyMetrologyTransportable2018}
J. Grotti, S. Koller, S. Vogt, S. H{\"a}fner, U. Sterr, C. Lisdat, H. Denker, C. Voigt, L. Timmen, A. Rolland, F.~N. Baynes, H.~S. Margolis, M. Zampaolo, P. Thoumany, M. Pizzocaro, B. Rauf, F. Bregolin, A. Tampellini, P. Barbieri, M. Zucco, G.~A. Costanzo, C. Clivati, F. Levi, and D. Calonico, Nature Physics {\bf 14},  437  (2018).

\bibitem{hachisuSItraceableMeasurementOptical2017a}
H. Hachisu, G. Petit, F. Nakagawa, Y. Hanado, and T. Ido, Optics Express {\bf 25},  8511  (2017).

\bibitem{kimImprovedAbsoluteFrequency2017a}
H. Kim, M.-S. Heo, W.-K. Lee, C.~Y. Park, H.-G. Hong, S.-W. Hwang, and D.-H. Yu, Japanese Journal of Applied Physics {\bf 56},  050302  (2017).

\bibitem{lemkeSpin122009}
N. Lemke, A. Ludlow, Z. Barber, T. Fortier, S. Diddams, Y. Jiang, S. Jefferts, T. Heavner, T. Parker, and C. Oates, Physical Review Letters {\bf 103},  063001  (2009).

\bibitem{lodewyckOpticalMicrowaveClock2016}
J. Lodewyck, S. Bilicki, E. Bookjans, J.-L. Robyr, C. Shi, G. Vallet, R. Le~Targat, D. Nicolodi, Y. Le~Coq, J. Gu{\'e}na, M. Abgrall, P. Rosenbusch, and S. Bize, Metrologia {\bf 53},  1123  (2016).

\bibitem{mcgrewOpticalSecondVerifying2019}
W.~F. McGrew, X. Zhang, H. Leopardi, R.~J. Fasano, D. Nicolodi, K. Beloy, J. Yao, J.~A. Sherman, S.~A. Sch{\"a}ffer, J. Savory, R.~C. Brown, S. R{\"o}misch, C.~W. Oates, T.~E. Parker, T.~M. Fortier, and A.~D. Ludlow, Optica {\bf 6},  448  (2019).

\bibitem{pizzocaroAbsoluteFrequencyMeasurement2017}
M. Pizzocaro, P. Thoumany, B. Rauf, F. Bregolin, G. Milani, C. Clivati, G.~A. Costanzo, F. Levi, and D. Calonico, Metrologia {\bf 54},  102  (2017).

\bibitem{rosenbandFrequencyRatioAl+and2008}
T. Rosenband, D.~B. Hume, P.~O. Schmidt, C.~W. Chou, A. Brusch, L. Lorini, W.~H. Oskay, R.~E. Drullinger, T.~M. Fortier, J.~E. Stalnaker, S.~A. Diddams, W.~C. Swann, N.~R. Newbury, W.~M. Itano, D.~J. Wineland, and J.~C. Bergquist, Science {\bf 319},  1808  (2008).

\bibitem{takamotoFrequencyRatiosSr2015}
M. Takamoto, I. Ushijima, M. Das, N. Nemitz, T. Ohkubo, K. Yamanaka, N. Ohmae, T. Takano, T. Akatsuka, A. Yamaguchi, and H. Katori, Comptes Rendus. Physique {\bf 16},  489  (2015).

\bibitem{fujiedaAdvancedSatelliteBasedFrequency2018}
M. Fujieda, S.-H. Yang, T. Gotoh, S.-W. Hwang, H. Hachisu, H. Kim, Y.~K. Lee, R. Tabuchi, T. Ido, W.-K. Lee, M.-S. Heo, C.~Y. Park, D.-H. Yu, and G. Petit, IEEE Transactions on Ultrasonics, Ferroelectrics, and Frequency Control {\bf 65},  973  (2018).

\bibitem{hisaiImprovedFrequencyRatio2021}
Y. Hisai, D. Akamatsu, T. Kobayashi, K. Hosaka, H. Inaba, F.-L. Hong, and M. Yasuda, Metrologia {\bf 58},  015008  (2021).

\bibitem{pizzocaroIntercontinentalComparisonOptical2021}
M. Pizzocaro, M. Sekido, K. Takefuji, H. Ujihara, H. Hachisu, N. Nemitz, M. Tsutsumi, T. Kondo, E. Kawai, R. Ichikawa, K. Namba, Y. Okamoto, R. Takahashi, J. Komuro, C. Clivati, F. Bregolin, P. Barbieri, A. Mura, E. Cantoni, G. Cerretto, F. Levi, G. Maccaferri, M. Roma, C. Bortolotti, M. Negusini, R. Ricci, G. Zacchiroli, J. Roda, J. Leute, G. Petit, F. Perini, D. Calonico, and T. Ido, Nature Physics {\bf 17},  223  (2021).

\bibitem{riehleCIPMListRecommended2018a}
F. Riehle, P. Gill, F. Arias, and L. Robertsson, Metrologia {\bf 55},  188  (2018).

\bibitem{NGSSurvey2019}
D. van Westrum, NOAA Technical Memorandum NOS NGS 77  (2019).

\bibitem{OhenhenLandSubsidence2025}
L.~O. Ohenhen, G. Zhai, J. Lucy, S. Werth, G. Carlson, M. Khorrami, F. Onyike, N. Sadhasivam, A. Tiwari, K. Ghobadi-Far, S.~F. Sherpa, J.-C. Lee, S. Zehsaz, and M. Shirzaei, Nature Cities {\bf 2},  543–554  (2025).

\bibitem{NGSSurvey2025}
D. van Westrum and K. Ahlgren, NOAA Technical Memorandum NOS NGS 99  (2025).

\bibitem{Note3}
Atomic transition frequency values for each clock are obtained by combining a measured beatnote $f_b$ against a comb tooth $n$ with the comb repetition rate $f_{rep}$, carrier envelope offset $f_0$, frequency multiplication factor $m$, and experimentally applied offsets $f_{os}$. $f_{rep}$ is set by the comb lock to the Si cavity; $f_0$ is controlled to a set value; and $m$ and $f_{os}$ are set experimentally by each clock. Together, the measured atomic transition frequency is $\nu =m\left (nf_{rep}+f_0+f_b\right )+f_{os}$ \cite {FortierCombReview2019}. We note that all RF sources are referenced to a UTC(NIST) traceable maser and measured by frequency counters referenced to the same maser. Because the absolute inaccuracy of the maser is small compared to the relevant optical frequencies and the counter resolution, we eliminate counter digitization noise by treating controlled RF frequencies as exact so long as the counted and set frequencies agree.

\bibitem{margolisGuidelinesEvaluationReporting2020a}
H. Margolis and M. Pizzocaro, Guidelines on the Evaluation and Reporting of Correlation Coefficients between Frequency Ratio Measurements, 2020.

\bibitem{DorscherYbSrRatio2021}
S. Dörscher, N. Huntemann, R. Schwarz, R. Lange, E. Benkler, B. Lipphardt, U. Sterr, E. Peik, and C. Lisdat, Metrologia {\bf 58},  015005  (2021).

\bibitem{mandel1970interlaboratory}
J. Mandel and R.~C. Paule, Analytical Chemistry {\bf 42},  1194  (1970).

\bibitem{langan2019comparison}
D. Langan, J.~P. Higgins, D. Jackson, J. Bowden, A.~A. Veroniki, E. Kontopantelis, W. Viechtbauer, and M. Simmonds, Research Synthesis Methods {\bf 10},  83  (2019).

\bibitem{merkatas2019shades}
C. Merkatas, B. Toman, A. Possolo, and S. Schlamminger, Metrologia {\bf 56},  054001  (2019).

\bibitem{possolo2023tracking}
A. Possolo, Statistical Science {\bf 38},  655  (2023).

\bibitem{kimEvaluationLatticeLight2023}
K. Kim, A. Aeppli, T. Bothwell, and J. Ye, Physical Review Letters {\bf 130},  113203  (2023).

\bibitem{bothwell2022}
T. Bothwell, C.~J. Kennedy, A. Aeppli, D. Kedar, J.~M. Robinson, E. Oelker, A. Staron, and J. Ye, Nature {\bf 602},  420  (2022).

\bibitem{lisdatBlackbodyRadiationShift2021}
{\relax Ch}. Lisdat, S. D{\"o}rscher, I. Nosske, and U. Sterr, Physical Review Research {\bf 3},  L042036  (2021).

\bibitem{blatt_Rabi_2009}
S. Blatt, J.~W. Thomsen, G.~K. Campbell, A.~D. Ludlow, M.~D. Swallows, M.~J. Martin, M.~M. Boyd, and J. Ye, Phys. Rev. A {\bf 80},  052703  (2009).

\bibitem{shi_polarizabilities_2015}
C. Shi, J.-L. Robyr, U. Eismann, M. Zawada, L. Lorini, R. Le~Targat, and J. Lodewyck, Phys. Rev. A {\bf 92},  012516  (2015).

\bibitem{ushijima_ls_2018}
I. Ushijima, M. Takamoto, and H. Katori, Phys. Rev. Lett. {\bf 121},  263202  (2018).

\bibitem{SchmidtQLS2005}
P.~O. Schmidt, T. Rosenband, C. Langer, W.~M. Itano, J.~C. Bergquist, and D.~J. Wineland, Science {\bf 309},  749  (2005).

\bibitem{HumeQLS2007}
D. Hume, T. Rosenband, and D. Wineland, Physical Review Letters {\bf 99},  120502  (2007).

\bibitem{KellerMMmeasurement2015}
J. Keller, H.~L. Partner, T. Burgermeister, and T.~E. Mehlstäubler, Journal of Applied Physics {\bf 118},  104501  (2015).

\bibitem{nemitz_2019}
N. Nemitz, A.~A. J\o{}rgensen, R. Yanagimoto, F. Bregolin, and H. Katori, Phys. Rev. A {\bf 99},  033424  (2019).

\bibitem{Bothwell_2025}
T. Bothwell, B.~D. Hunt, J.~L. Siegel, Y.~S. Hassan, T. Grogan, T. Kobayashi, K. Gibble, S.~G. Porsev, M.~S. Safronova, R.~C. Brown, K. Beloy, and A.~D. Ludlow, Phys. Rev. Lett. {\bf 134},  033201  (2025).

\bibitem{Chen24}
C.-C. Chen, J.~L. Siegel, B.~D. Hunt, T. Grogan, Y.~S. Hassan, K. Beloy, K. Gibble, R.~C. Brown, and A.~D. Ludlow, Phys. Rev. Lett. {\bf 133},  053401  (2024).

\bibitem{hassan_2025}
Y.~S. Hassan, K. Beloy, J.~L. Siegel, T. Kobayashi, E. Swiler, T. Grogan, R.~C. Brown, T. Rojo, T. Bothwell, B.~D. Hunt, A. Halaoui, , and A.~D. Ludlow, Phys. Rev. Lett. {\bf 135},  063402  (2025).

\bibitem{beloy_2014}
K. Beloy, N. Hinkley, N.~B. Phillips, J.~A. Sherman, M. Schioppo, J. Lehman, A. Feldman, L.~M. Hanssen, C.~W. Oates, and A.~D. Ludlow, Phys. Rev. Lett. {\bf 113},  260801  (2014).

\end{thebibliography}
